\documentclass[pdflatex, sn-nature, iicol]{sn-jnl}
\pdfoutput=1
\usepackage{amsmath,amssymb,amsfonts}
\usepackage{algorithm}
\usepackage{dsfont}
\usepackage{physics}
\usepackage{mathtools}
\usepackage[separate-uncertainty = true,multi-part-units=single]{siunitx}
\usepackage[all]{hypcap}
\usepackage{pdflscape}
\usepackage{xcolor}

\usepackage[sort&compress,numbers]{natbib}
\usepackage{doi}
\newcommand\mybar{\kern-2pt\rule[-.8pt]{.8pt}{7pt}\kern1pt}

\makeatletter

\let\corremailfootnote\@empty

\renewcommand{\email}[1]{%
  \global\advance\emailcnt by 1\relax
  \if@corauemail
    \ifx\corremailfootnote\@empty
      \gdef\corremailfootnote{%
        \href{mailto:#1}{\nolinkurl{#1}}%
      }%
    \else
      \g@addto@macro\corremailfootnote{%
        ; \href{mailto:#1}{\nolinkurl{#1}}%
      }%
    \fi
  \else
    \g@addto@macro\authemail{%
      \setcounter{footnote}{0}%
      \textcolor{blue}{#1};\ %
    }%
  \fi
}

\renewcommand{\@@corrauthor}[2][]{%
  \def\@authfrstarg{#1}%
  \@corauemailtrue
  \advance\corraucount by 1%
  \g@addto@macro\artauthors{%
    \global\@auemailtrue
    \Authorfont
    \def\baselinestretch{1}%
    \authorsep{#2}\unskip
    \ifx\@authfrstarg\empty
      \textsuperscript{*}%
    \else
      \textsuperscript{\smash{{%
        \@for\@@affmark:=#1\do{%
          \edef\affnum{%
            \@ifundefined{X@\@@affmark}{\@@affmark}{\jmkRef{\@@affmark}}%
          }%
          \unskip\sep\affnum\let\sep=,%
        }%
        \unskip,*%
      }}}%
    \fi
    \unskip
    \def\authorsep{\au@and~}%
    \global\let\sep\@empty
    \global\let\@corref\@empty
  }%
}

\newcommand{\printcorremailfootnote}{%
  \ifx\corremailfootnote\@empty
  \else
    \begingroup
      \renewcommand{\thefootnote}{*}%
      \footnotetext[1]{\corremailfootnote}%
    \endgroup
  \fi
}

\makeatother

\begin{document}
\title{Scalable Single-Step Generation of W States in 2D Superconducting Qubit Lattices}

\author*[1,2]{J.~H.~Romeiro}\email{joao.romeiro@wmi.badw.de}
\author[2,3]{F.~A.~Roy}
\author[1,2]{N.~Bruckmoser}
\author[1,2]{I.~Tsitsilin}
\author[1,2]{N.~J.~Glaser}
\author[1,2]{C.~M.~F.~Schneider}
\author[1,2]{G.~B.~P.~Huber}
\author[1,2]{S.~A.~Sch\"obe}
\author[1,2]{J.~Schirk}
\author[1,2]{F.~Wallner}
\author[1,2]{M.~Singh}
\author[1,2]{J.~Feigl}
\author[1,2]{L.~Koch}
\author[1,2]{L.~Södergren}
\author[1,2]{M.~Werninghaus}
\author[1,2,4]{S.~Filipp}

\affil[1]{Technical University of Munich, TUM School of Natural Sciences, Department of Physics, Garching 85748, Germany}
\affil[2]{Walther-Meißner-Institut, Bayerische Akademie der Wissenschaften, 85748 Garching, Germany}
\affil[3]{Theoretical Physics, Saarland University, 66123 Saarbrücken, Germany}
\affil[4]{Munich Center for Quantum Science and Technology (MCQST), Schellingstraße 4, 80799 München, Germany}

\abstract{
The reliable generation of multi-qubit entanglement is a prerequisite for large-scale quantum information technologies. 
In particular, W~states are a valuable resource owing to their resilience under local loss or measurement.
Nevertheless, preparing these states with sequential two-qubit gates often requires substantial time overhead.
By contrast, engineered simultaneous interactions enable fast entanglement generation, even in qubit systems with limited nearest-neighbour connectivity.
Here, we demonstrate a set of fast and robust operations for coherently distributing a single excitation across a lattice of arbitrary size, thereby directly generating W~states from initial product states. 
In 2D lattices, the excitation propagates along both directions simultaneously, such that the total entanglement time scales only with the largest dimension. We exploit this property to prepare a six-qubit W~state in a 3$\times$2 superconducting lattice within $\SI{99}{\ns}$, achieving a tomographic fidelity of $\SI{83.9\pm 1.0}{\%}$.
We then extend the protocol to create entanglement across chains of up to seven qubits, with the largest W~state generated in $\SI{264}{\ns}$ with a fidelity of $\SI{79.6 \pm 1.3}{\%}$.
}
\maketitle
\printcorremailfootnote
\begin{figure*}[t!]
\centering
\includegraphics{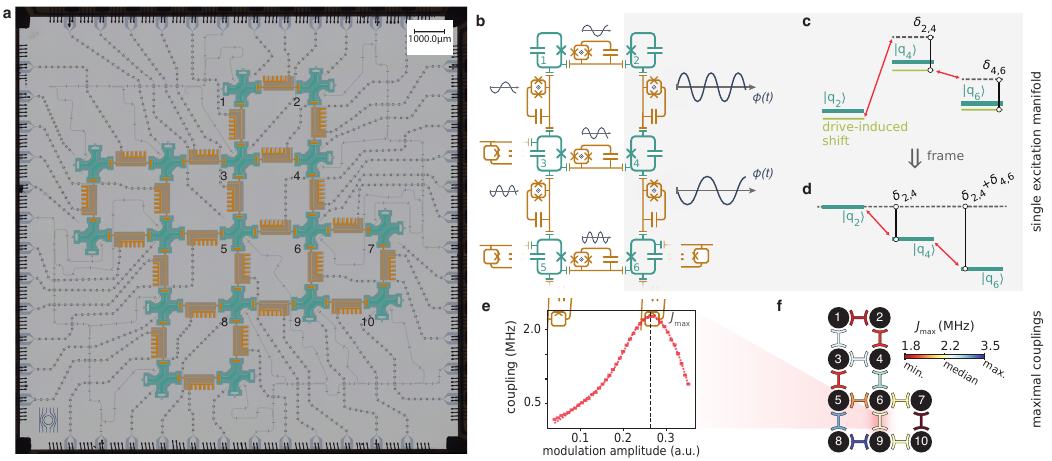}
\caption{
    \label{fig:overview}
\textbf{Superconducting qubit device with parametrically-driven qubit chains.}
\textbf{a} False-coloured image of the processor, containing a total of 17 transmon qubits (cyan) and 24 tunable couplers (amber).
Numbers indicate the subset of 10 qubits used in this work.
\textbf{b}~Circuit diagram showing the operation of a $3\times2$ register comprising qubits $\text{q}_1$ through $\text{q}_6$.
Simultaneous AC flux modulations (insets) are applied near the difference frequency of the corresponding qubit pair.
\textbf{c} Energy diagram of the single-excitation manifold, showing effective couplings (red) between qubits $\text{q}_2$, $\text{q}_4$ and $\text{q}_6$.
Parametric drives induce AC-Stark shifts (lime green) of the qubit levels (cyan).
Black vertical lines indicate the detunings $\delta_{2,4}$ and $\delta_{4,6}$ between each drive frequency and its corresponding transition.
All detunings and shifts are exaggerated for clarity, and couplings to other qubits are omitted.
\textbf{d} Equivalent energy diagram in the rotating frame. 
The energy difference between adjacent levels is set by the respective drive detuning, $\delta_{2,4}$ or $\delta_{4,6}$.
\textbf{e} Experimental coupling strength between qubits $\text{q}_{6}$ and $\text{q}_{9}$ as a function of the drive amplitude, exhibiting a maximal value of $J_{\text{max}}/2\pi\approx \SI{2.25}{\mega\hertz}$ (vertical dashed line).
\textbf{f} Heat~map summarising the maximal couplings achieved across all used pairs, ranging from $\SI{1.8}{\mega\hertz}$ (qubits $\text{q}_{7}$ and $\text{q}_{10}$) to $\SI{3.5}{\mega\hertz}$ (qubits $\text{q}_{8}$ and $\text{q}_{9}$). 
The median value for all couplers is $\SI{2.2}{\mega\hertz}$.    
    }
\end{figure*}

\section*{Introduction}
The generation of multipartite entanglement is a cornerstone of quantum information technology, providing the basis for key tasks in quantum error correction~\cite{Raussendorf2007_surfacecode, Shor1995_QEC, Brun2006_EAQECC}, distributed~\cite{Cirac1999_DQC, Main2025_DQCphotonic} and measurement-based quantum computing~\cite{Raussendorf2001_OneWayMBQC, Briegel2009_MBQC, Li2023_multipartiteMBQC}, quantum communication~\cite{Zhao2004_opendestination, Cleve1999_QSS, Bell2014_QSSphotonic}, and quantum sensing~\cite{toth2012_multipartitemetrology}. 
For any system with more than two qubits, multiple inequivalent classes of entangled states exist, exhibiting distinct properties and applications.
W~states stand out due to their resilience under local loss or measurement, as reading out one qubit can leave the remaining qubits entangled with a probability approaching unity for large systems~\cite{Dur2000_Wstate, Briegel2001_persistency}.
W~states are highly attractive for distributed network storage~\cite{choi2010_entangledmemories, miguel2020_dickedelocalized}, secure communication~\cite{joo2002_wsecure, liu2011_wcomparison, lipinska2018_wanonymous}, state teleportation~\cite{shi2002_teleportation, agrawal2006_wteleportation} and distillation~\cite{fortescue2007_distillation, greenwood2025_Wdistillationscq}, magnetometry~\cite{ng2014_wmagneticgradient}, and others. 
To meet practical demands, these applications require the reliable generation of large entangled states, an endeavour that remains highly challenging. 

A W~state is defined as
\begin{equation}
    \ket{\text{W}_N}\coloneqq\frac{1}{\sqrt{N}}\Big(\ket{10\ldots0}+\ket{01\ldots0}\ldots+\ket{00\ldots1}\Big),
    \label{eq:defW}
\end{equation}
thus describing a single excitation coherently delocalised across $N$ qubits.
While W~states can be generated via sequential two-qubit gates~\cite{Baertschi2022_shortcircuit, Baertschi2022_divideconquer, hu2024_effQST, chen2025_efficient2q}, this approach can lead to large circuit depths in platforms with limited connectivity, such as superconducting qubits.
Alternatively, simultaneous nearest-neighbour couplings allow multiple qubits to interact in a single step, thus reducing the total operation time.
This method has been successfully demonstrated in superconducting platforms for a variety of applications, including single-step GHZ state preparation~\cite{Song2017_10qubit, RoyRomeiro2025_GHZ}, state transfer~\cite{Xiang2024_2Dstatetransfer, liu2025_thouless, wang2025_zigzag}, analog simulation~\cite{Karamlou2022_transport, 
Rosen2024_vectorpotential, rosen2024_flat, wang2026_whichpath}, and multi-qubit gates~\cite{Warren2023_3qgates}.
In this context, several protocols for preparing W~states make use of simultaneous interactions with a common mode~\cite{hume2009_dickeions, kang2016_fastw, Wang2020_radiant, zhang2023_wqedw}.
Nevertheless, the underlying requirement for all-to-one connectivity limits scalability towards larger qubit numbers.

In addition, single-step W~state generation has been investigated in 1D arrays, both in spin~\cite{wang2001_xyw, Kay2010_PST, Kay2017_generatestates, Kay2017_tailoring} and photonic systems~\cite{Perezleija2013_multiportw, Vildoso2023_lowlossWGAs, bugarski2024_waveguidew}.
These methods require engineering local interactions to rapidly distribute an excitation across all qubits, which imposes specific constraints on the spectrum and eigenstates of the Hamiltonian.
Specifically, determining a Hamiltonian satisfying these conditions constitutes an inverse eigenvalue problem \cite{Kay2017_generatestates, petrovic2018_opticaltransfer, wang2025_inverse}, which typically admits multiple solutions.

Here, we extend these methods to demonstrate single-step generation of large W~states across a 2D superconducting qubit lattice.
In particular, we propose a novel set of time-independent Hamiltonians that evolve a local excitation into a uniform superposition after precisely half a period of the dynamics.
Intriguingly, our method allows the excitation to propagate across multiple spatial directions simultaneously, such that the total entanglement time in an $L\times M$ lattice is determined only by the largest dimension, $\mathcal{O}(\max(L,\:M))$, whereas sequential two-qubit gates require a circuit depth of $\mathcal{O}(L+M)$.
Using simultaneous parametric drives to control local couplings, we realise this protocol to show genuine single-step entanglement in a scalable transmon architecture, with up to six qubits in 2D and seven qubits in 1D.

\section*{Results}
\subsection*{Device Description}
\label{sec:device_description}
We perform experiments on the superconducting device shown in Fig.~\ref{fig:overview}\textbf{a}, where the numbered qubits indicate the subset used in this work.
All qubits possess dedicated XY control lines and resonators for dispersive readout. 
Resonators are grouped into four distinct feedlines compatible with multiplexed readout. 
Qubits are arranged in a 2D square lattice, where interactions between each neighbouring pair $\{\text{q}_i, \text{q}_j\}$ are mediated by a tunable coupler, $\text{c}_{ij}$. 
The Hamiltonian describing all non-linear elements is 
\begin{equation}
	\begin{split}
	\hat{H}_{\text{full}}/\hbar = 
    &\sum_{i} \Big[\omega_{\text{q}_i}\hat{a}^\dagger_{\text{q}_i} \hat{a}_{\text{q}_i} + \frac{\alpha_{\text{q}_i}}{2}\hat{a}^\dagger_{\text{q}_i} \hat{a}_{\text{q}_i}^\dagger\hat{a}_{\text{q}_i} \hat{a}_{\text{q}_i}\Big]
    \\ 
    + \sum_{\{i,j\}}&\Big[\omega_{\text{c}_{ij}}(\phi_{\text{c}_{ij}})\hat{a}_{\text{c}_{ij}}^\dagger\hat{a}_{\text{c}_{ij}} + \frac{\alpha_{\text{c}_{ij}}}{2}\hat{a}^\dagger_{\text{c}_{ij}} \hat{a}^\dagger_{\text{c}_{ij}}\hat{a}_{\text{c}_{ij}} \hat{a}_{\text{c}_{ij}}\Big]\\
    +\sum_{\{i,j\}}&\Big[\frac{g_{ij}}{2}(\hat{a}_{\text{q}_i}^\dagger-\hat{a}_{\text{q}_i})(\hat{a}_{\text{c}_{ij}}^\dagger-\hat{a}_{\text{c}_{ij}})\\
    &\qquad\qquad+ \frac{g_{ji}}{2}(\hat{a}_{\text{q}_j}^\dagger-\hat{a}_{\text{q}_j})(\hat{a}_{\text{c}_{ij}}^\dagger-\hat{a}_{\text{c}_{ij}})\Big],
    \end{split}
    \label{eq:Hamiltonian_couplers}
\end{equation}
where the operators $\hat{a}^{\dagger}$ and $\hat{a}$ denote the raising and lowering operators of the sub-indexed mode, with frequency~$\omega$ and anharmonicity~$\alpha$. 
The static coupling between coupler $\text{c}_{ij}$ and qubit $\text{q}_i$ ($\text{q}_j$) is denoted by $g_{ij}$ ($g_{ji}$).
Coupler frequencies are tuned by local magnetic fields~$\phi_{c_{ij}}$, which are controlled via dedicated flux lines.
Additional information on the experimental setup is provided in \ref{app_setup}.

During operation, all couplers are DC flux-biased in the dispersive regime relative to the qubits, i.e. $\omega_{\text{c}_{ij}}(\phi_{c_{ij}})-\omega_{\text{q}_{i/j}}\gg g_{ij/ji}$. 
To activate qubit interactions, AC flux modulations are simultaneously applied to selected couplers, as illustrated in Fig.~\ref{fig:overview}\textbf{b}.
Each modulation frequency $\omega_{\phi_{ij}}$ is set near the corresponding qubit-qubit detuning $\omega_{\text{q}_j}-\omega_{\text{q}_i}$, thereby inducing parametric exchange-type interactions within the single-excitation manifold, as shown in Fig.~\ref{fig:overview}\textbf{c}~\cite{Mckay2016_tunablebus, Ganzhorn2020_benchmark}. 
Indeed, a time-dependent Schrieffer--Wolff transformation eliminates the couplers from Eq.~(\ref{eq:Hamiltonian_couplers})~\cite{Roth2017_exchangetype}, resulting in the effective Hamiltonian in the single-excitation manifold 
\begin{equation}
\begin{split}
\hat{H}_{\text{eff}}/\hbar = \sum_{i\in V} &\tilde{\omega}_{\text{q}_i}\ket{\text{q}_i}\bra{\text{q}_i} \\ 
+ \sum_{\{i,j\}\in E}&\Omega_{ij}\big(\phi_{c_{ij}},\:t\big)|\text{q}_j\rangle\bra{\text{q}_i} + \text{h.c.},
\end{split}
\label{eq:SW}
\end{equation} 
where $V$ indicates the set of interacting qubits and
$E$ the set of modulated couplers.
The state \mbox{$\ket{\text{q}_i}\coloneqq\ket{0\ldots1_i\ldots0}$} denotes an excitation localised on qubit $\text{q}_i$, with its frequency $\tilde{\omega}_{\text{q}_i}$ adapted to include drive-induced shifts. 

Importantly, each coupling coefficient $\Omega_{ij}(\phi_{c_{ij}},\:t)$ in Eq.~(\ref{eq:SW}) contains a Fourier component $\propto e^{- i\omega_{\phi_{ij}}t}$, originating from the corresponding AC flux modulation~\cite{Roth2017_exchangetype}.
By moving to a suitable rotating frame (see \hyperref[sec:met1]{Methods section}), this component becomes stationary for all interacting pairs.
After a rotating-wave approximation, the resulting Hamiltonian takes the time-independent form 
\begin{equation}
\begin{split}
\hat{H}/\hbar \approx \sum_{i\in V}& \Delta_{i}\ket{\text{q}_i}\bra{\text{q}_i} \\ 
+ \sum_{\{i,j\}\in E}&(J_{ij}|\text{q}_j\rangle\bra{\text{q}_i} + \text{h.c.}).
\end{split}
\label{eq:eff}
\end{equation}
In this frame, on-site energies $\Delta_i$ are controlled by the drive detunings $\delta_{ij}:=\omega_{\phi_{ij}}-(\tilde{\omega}_{\text{q}_j}-\tilde{\omega}_{\text{q}_i})$, such that \mbox{$\Delta_i-\Delta_j=\delta_{ij}$} for all interacting pairs, as illustrated in Fig.~\ref{fig:overview}\textbf{d}.
Meanwhile, the effective coupling strengths $|J_{ij}|$ are set by the respective modulation amplitude $A_{ij}$, according to
\begin{equation}
  |J_{ij}| \propto g_{ij} g_{ji}
  \, \mathcal{J}_1 \!\left(
  \left.\frac{\partial \omega_{c_{ij}}}{\partial \phi_{c_{ij}}}\right|_{\phi^{\text{DC}}}
  \frac{A_{ij}}{\omega_{\phi_{ij}}}
  \right)\!,
\label{eq:bessel}
\end{equation}
where $\mathcal{J}_1$ is a Bessel function of the first kind~\cite{Mckay2016_tunablebus, Huber2025_parametric}.
In practice, the maximum achievable couplings are further restricted by strong-drive effects such as leakage to other transitions and qubit ionization effects~\cite{xia2025_parametricionize}.
An example of coupling tunability as a function of drive amplitude is shown in Fig.~\ref{fig:overview}\textbf{e}, and the maximal values measured for all thirteen used couplers are summarised in~Fig.~\ref{fig:overview}\textbf{f}.

To directly generate a W~state across all $N\coloneqq|V|$ qubits, we first determine the required couplings $J_{ij}$ and on-site energies $\Delta_{i}$ by solving the inverse eigenvalue problem
\begin{equation}
    e^{-i\hat{H}\tau/\hbar}\ket{\text{q}_\text{centre}} = \ket{\text{W}_{N}}\!,
\label{eq:Wstatecondition}
\end{equation}
where the synthesis time $\tau>0$ is chosen such that all~$|J_{ij}|$ are within the experimental range.
The excitation is initially localised on qubit $\text{q}_\text{centre}$, which is selected near the geometric centre to minimise propagation distance. 
We then implement the target Hamiltonian $\hat{H}$ experimentally, with the amplitudes and frequencies of AC flux modulations determined as follows.
During calibration, the excitation is prepared on~$\text{q}_\text{centre}$ using a $\pi$~pulse, after which the flux drives are simultaneously applied.
This sequence is repeated for multiple drive durations, each followed by measurement of all $N$~qubits, and the resulting populations are compared with those predicted by $\hat{H}$. 
The discrepancy is used as a cost function for a closed-loop CMA-ES optimiser~\cite{Akiba2019_optuna},
which iteratively updates the drive parameters until convergence~\cite{werninghaus2021_leakage, glaser2025_closedloop}.
Following the calibration procedure, the state at time $\tau$ corresponds to $\ket{\text{W}_N}$ up to relative phases between the basis states $\ket{\text{q}_i}$, which are subsequently corrected using single-qubit virtual Z rotations~\cite{Mckay2017_virtualz}.

\subsection*{Three-qubit chains}
\begin{figure}[b]
\centering
\includegraphics{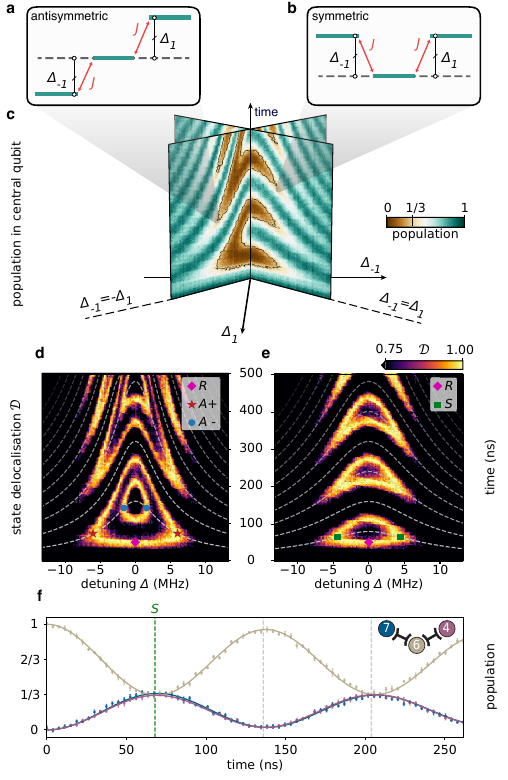}
\caption{
    \label{fig:3Q}
\textbf{Single-step W~state generation in three-qubit chains.}
Diagrams show the single-excitation energy levels (cyan) with uniform couplings $J$ (red).
Detunings $\Delta_{\pm1}$ are equal in magnitude and follow either an \textbf{a} antisymmetric or \textbf{b} symmetric profile.
\textbf{c} Population of the central qubit, initially prepared in its excited state,  as a function of detuning magnitude. Contours highlight the parameter regions in which the central population is exactly one third.
The state delocalisation $\mathcal{D}$ is shown for \textbf{d} antisymmetric and \textbf{e} symmetric Hamiltonians, with $\mathcal{D}=1$ corresponding to a W~state.
The fastest W~state generation occurs at resonance~$R$ (magenta diamonds).
Dashed lines indicate integer multiples of the half-period. 
The detuning choices $A_+$ (red stars), $A_-$ (blue~circles) and $S$ (green squares) prepare W~states at exactly one half-period.
\textbf{f} Population dynamics for the symmetric Hamiltonian ($S$) using qubits $\text{q}_{7}$, $\text{q}_{6}$ and $\text{q}_{4}$, with a synthesis time of $\tau=\SI{68}{\ns}$ (green dashed line).
Fits (solid lines) are obtained from Eq.~(\ref{eq:H_spin_3}) with additional Lindbladian terms accounting for qubit dephasing and excitation loss.
}
\end{figure}

To illustrate the different Hamiltonian structures satisfying Eq.~(\ref{eq:Wstatecondition}), we first generate W~states in a three-qubit chain comprising $\text{q}_{8}$, $\text{q}_{9}$ and $\text{q}_{10}$.
Specifically, we consider settings where an excitation prepared on the central qubit,~$\text{q}_{9}$, spreads to the two outer qubits with equal populations.
In the localised basis~$\{\ket{\text{q}_{8}},\ket{\text{q}_{9}},\ket{\text{q}_{10}}\}$, each implemented Hamiltonian takes the form
\begin{equation}
\hat{H}_3/\hbar =
\begin{pmatrix}
\Delta_{-1} & J^* & 0 \\
J & 0 & J^* \\
0 & J & \Delta_{1}
\end{pmatrix}\!,
\label{eq:H_spin_3}
\end{equation}
with $\Delta_{-1}=\mp\Delta_1$, corresponding to the antisymmetric and symmetric cases shown in Fig.~\ref{fig:3Q}\textbf{a},\textbf{b}.

To realise $\hat{H}_3$ experimentally, we first calibrate the two parametric drives to be resonant with their respective transitions and to yield identical coupling strengths \mbox{$|J|/2\pi\approx\SI{2.21}{\mega\hertz}$}.
We then detune both drives equally, either with opposite (\mbox{$\Delta_{-1}=-\Delta_1$}) or equal signs (\mbox{$\Delta_{-1}=\Delta_1$}) relative to the central qubit.
The time evolution of an excitation initialised on qubit $\text{q}_{9}$ is recorded for varying detuning magnitudes \mbox{$\Delta\coloneqq|\Delta_{\pm1}|$}, as shown in Fig.~\ref{fig:3Q}\textbf{c}.
The period $T$, defined as the time after which the excitation refocuses on the central qubit, follows from diagonalising Eq.~(\ref{eq:H_spin_3}) as   
\begin{equation}
    T(\Delta) =
    \frac{2\pi}{\sqrt{\Delta^2 + 2k|J|^2}},
    \label{eq:period}
\end{equation}
with $k=1$ and $k=4$ for antisymmetric and symmetric Hamiltonians, respectively.
In both cases, 
the population on $\text{q}_9$ reaches one third at a detuning-dependent fraction of the period. 
At these times, the excitation is equally distributed across the chain, as quantified by the state delocalisation 
\mbox{$\mathcal{D}\coloneqq\big(3(P^2_{8} + P^2_{9} + P^2_{10})\big)^{-1}$}, calculated from the qubit populations $P_{i}$ and shown in Fig.~\ref{fig:3Q}\textbf{d},\textbf{e}.
In particular, $\mathcal{D}=1$ indicates a W~state.

We now identify the parameter choices most relevant to our protocol.
First, the fastest W~state generation occurs in the resonant case with $\Delta=0$, for which $\mathcal{D}=1$ at $\tau_R = \arctan(\sqrt{2})/(|J|\sqrt{2})$.
Alternatively, a W~state is generated at precisely one half-period ($T/2$) for two antisymmetric detuning choices,
\begin{equation}
|\Delta_{A\pm}| = (1\pm\sqrt{3})\,|J|,
\label{eq:dets_3Q_anti}
\end{equation}
and one symmetric choice,
\begin{equation}
|\Delta_S| = 2\,|J|\,. 
\label{eq:dets_3Q_sim}
\end{equation}
These half-period solutions can provide enhanced robustness against certain parameter fluctuations, as the first-order time derivative vanishes at the synthesis time $\tau:=T/2$ (see~\ref{app_robustness}).
Notably, the symmetric choice $\Delta_S$ exhibits a shorter period, with \mbox{$T(\Delta_S)/T(\Delta_{A+})\approx0.888$} according to Eq.~(\ref{eq:period}).

We therefore realise the symmetric half-period operation using qubits $\text{q}_{7}$, $\text{q}_{6}$ and $\text{q}_{4}$, with the resulting dynamics highlighted in
Fig.~\ref{fig:3Q}\textbf{f}.
The chosen synthesis time is $\tau=\SI{68}{ns}$, corresponding to \mbox{$|J|/2\pi\approx\SI{2.13}{\mega\hertz}$}.
The populations are fitted to the dynamics generated by the Hamiltonian in Eq.~(\ref{eq:H_spin_3}), with additional Lindbladian terms accounting for both dephasing and excitation loss on all qubits. 
We reconstruct the three-qubit state at time $\tau$ via Quantum State Tomography (QST)~\cite{Hradil1997_QSTMLE, Steffen2006_QST}, by measuring all $3^3$ local Pauli operators. 
After applying assignment-matrix inversion to mitigate readout errors in the tomographic data~\cite{Bravyi2021_REM}, we obtain a W~state fidelity of \mbox{$F_\text{W}=\SI{94.3\pm0.3}{\%}$}, averaged over multiple QST repetitions.
The dominant error source consists of population assigned to other excitation-number manifolds, \mbox{$\epsilon_{\neq1}\approx\SI{4.17}{\%}$}, which is primarily attributed to unmitigated State Preparation and Measurement (SPAM) errors, including the single-qubit rotations used for tomography. 
Moreover, dephasing and excitation losses contribute \mbox{$\epsilon_{\text{dec.}}\approx\SI{0.86}{\%}$} to the infidelity, as estimated from the fitted population dynamics. 

\subsection*{Square qubit lattices}
In higher-dimensional systems, entanglement can be generated along multiple spatial directions in a genuinely parallel manner.
Indeed, by combining several 1D operations to form a square lattice, the time required for state preparation scales with one of the dimensions only. 
This is in stark contrast to approaches based on sequential two-qubit gates, where entangling any qubit with more than a single neighbour per step is prohibited.

\begin{figure}[t!]
\centering
\includegraphics{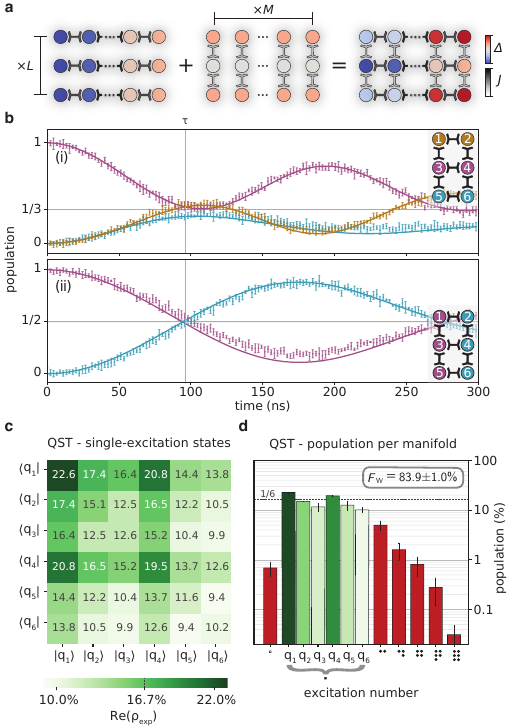}
\caption{
    \label{fig:2D}
    \textbf{Single-step W~state generation in 2D lattices.} \textbf{a}~Construction of a $L\times M$ square lattice from two overlapping sets of 1D Hamiltonians, resulting in independent single-excitation dynamics along each direction.
    \textbf{b} Single-step generation of a $3\times2$ W~state, with qubit $\text{q}_{4}$ initially excited. Dynamics are shown separately for rows (i) and columns (ii), with populations summed over the perpendicular direction. At the common synthesis time $\tau=\SI{99}{ns}$ (grey vertical line), the excitation is maximally delocalised along both axes (grey horizontal line). Fits (solid lines) include qubit dephasing and excitation loss.
    \textbf{c} Matrix elements of the reconstructed state $\bra{\text{q}_{i}}\rho_{\text{exp}}\ket{\text{q}_{j}}$, showing correlations between all single-excitation states.
    Only the real parts relevant to the fidelity estimate are shown; for clarity, all elements are expressed as percentages.
    \textbf{d} Populations of the reconstructed state $\rho_{\text{exp}}$.
    Single-excitation states are shown separately (green bars), with the target population of $1/6$ indicated by the horizontal dashed line.
    All remaining states are grouped by excitation-number manifolds (red bars).
    In total, these account for approximately $\SI{8.4}{\%}$ of the recovered population, with a clear suppression at larger excitation numbers.
    }
\end{figure}

We consider a 2D square lattice with \mbox{$N=L\times M$} qubits and nearest-neighbour couplings along each axis.
The single-excitation manifold is spanned by the basis of fully localised states $\left\{\ket{\text{q}_{l,m}}\right\}$, where \mbox{$l=1,\ldots,L$} and \mbox{$m=1,\ldots,M$} indicate the row and column coordinates, respectively.
We choose the nearest-neighbour interactions such that each row (column) realises a copy of the same 1D Hamiltonian up to a global offset, as shown in Fig.~\ref{fig:2D}\textbf{a}.
Under this condition, the effective Hamiltonian is expressed as two commuting terms,
\begin{equation}
    \hat{H}_{LM}=\hat{H}_L\otimes \hat{\mathbb I}_M+\hat{\mathbb I}_L\otimes \hat{H}_M,
\label{eq:2D_decomposition}
\end{equation}
each operating non-trivially along one direction only.
As a result, $\hat{H}_{LM}$ is a 2D solution to Eq.~(\ref{eq:Wstatecondition}) if and only if $\hat{H}_L$ and $\hat{H}_M$ are 1D solutions with commensurable synthesis times.
We emphasise that Eq.~(\ref{eq:2D_decomposition}) is limited to single-excitation dynamics, since in higher excitation-number manifolds the two subsystems effectively interact due to the double-occupancy constraint of hard-core bosons.

To demonstrate the 2D protocol, we experimentally generate a $3\times 2$ W~state using qubits $\text{q}_1$ through $\text{q}_6$. 
The synthesis time $\tau=\SI{99}{\ns}$ is set by the achievable interaction strengths along the largest dimension $M=3$, where all couplings are fixed at \mbox{$|J_M|/2\pi\approx\SI{1.46}{\mega\hertz}$} and drive detunings are set to $2|J_M|$, following the symmetric Hamiltonian derived in Eq.~(\ref{eq:dets_3Q_sim}).
At the same time, the shorter dimension $L=2$ evolves with \mbox{$|J_L|/2\pi=1/8\tau\approx\SI{1.26}{\mega\hertz}$} and zero detunings, which corresponds to a standard $\sqrt{i\text{SWAP}}$ operation of the same duration. 
When $\text{q}_4$ is initialised, the row and column populations reproduce the targeted three- and two-qubit dynamics respectively, resulting in a fully delocalised excitation as shown in Fig.~\ref{fig:2D}\textbf{b}.

Importantly, excitation dynamics across any closed loop are sensitive to the parametric drive phases, which determine the complex argument $\varphi$ of each coupling~\mbox{$J=|J|e^{i\varphi}$}. 
Indeed, alternative excitation paths with differing accumulated phases produce an interference pattern analogous to the Aharonov-Bohm effect~\cite{Rosen2024_vectorpotential}, as demonstrated in \ref{app_AB}.
To obtain the desired dynamics described by Eq.~(\ref{eq:2D_decomposition}), the total phase accumulated around every closed loop must be an integer multiple of 2$\pi$, thus ensuring constructive interference between the different trajectories.
This is enforced in the $3\times 2$ lattice by adjusting two parametric drive phases, corresponding to qubit pairs $(\text{q}_{2}, \:\text{q}_{4})$ and $(\text{q}_{4}, \:\text{q}_{6})$, while all others remain fixed.

We benchmark the W~state generation by reconstructing the full six-qubit density matrix $\rho_{\text{exp}}$ at time $\tau$, obtaining an average state fidelity of \mbox{$F_{\text{W}}=\SI{83.9\pm1.0}{\%}$} over multiple QST repetitions.
The main error source is qubit dephasing during the operation, estimated from the fitted population dynamics as $\epsilon_{\text{deph.}}\approx \SI{8.7}{\%}$.
This contribution is reflected in the single-excitation matrix elements $\bra{\text{q}_i}\rho_{\text{exp}}\ket{\text{q}_j}$, shown in Fig.~\ref{fig:2D}\textbf{c}.
In addition, populations assigned to higher excitation-number manifolds contribute a tomographic error of $\epsilon_{>1}\approx \SI{7.7}{\%}$, as summarised in Fig.~\ref{fig:2D}\textbf{d}.
These assignments are primarily due to unmitigated SPAM errors and crosstalk affecting the pre-tomographic gates (see \hyperref[sec:dis]{Discussion section}). 

\subsection*{Extended qubit chains and scaling}

\begin{figure}[b!]
\centering
\includegraphics{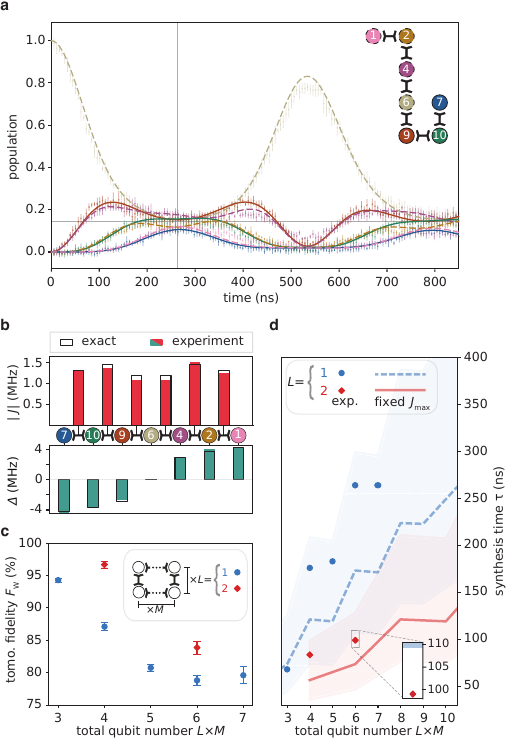}
\caption{
    \label{fig:scaling}
    \textbf{W~state generation in extended chains and scaling.}
    \textbf{a}~Single-step entanglement in a seven-qubit chain, with the central qubit $\text{q}_{6}$ initialised in its excited state.
    The excitation spreads symmetrically across the chain, resulting in equivalent dynamics for qubit pairs equidistant from the centre.
    For visual clarity, fits are shown as dashed lines for the first four qubits and solid lines for the remaining three.
    \textbf{b} Fitted Hamiltonian parameters (solid bars) compared with the exact solution (wireframes).
    On-site energies (cyan) exhibit an antisymmetric distribution with respect to the centre of the chain, while couplings (red) are symmetric in magnitude.
    \textbf{c} Average state fidelities extracted from QST, for all executed 1D ($L=1$, blue circles) and 2D operations ($L=2$, red diamonds).
    \textbf{d} Synthesis time for lattices with one dimension fixed to $L=1$ (blue) or $L=2$ (red).
    Solid points indicate realised operations, while solid lines show the expected scaling assuming a fixed maximal coupling of $J_{\text{max}}/2\pi=\SI{2.2}{\mega\hertz}$ (median value across all couplers).
    Shaded areas indicate the synthesis times accessible over the coupling range in our device ($\SIrange{1.8}{3.5}{\mega\hertz}$).
    The inset highlights the speedup obtained by the $3\times 2$ implementation compared to any feasible 1D operation. 
    }
\end{figure}

To extend our protocol to chains of \mbox{$M>3$} qubits, we introduce a parametrised family of Hamiltonians for numerically solving the inverse eigenvalue problem in Eq.~(\ref{eq:Wstatecondition}).
While previous work has derived solutions in odd-sized 1D chains with strictly resonant couplings~\cite{Perezleija2013_multiportw, Vildoso2023_lowlossWGAs, bugarski2024_waveguidew}, our construction extends these results to even lengths and includes non-zero detunings \mbox{$\Delta_j\neq\Delta_i$} to enable robust generation at exactly one half-period.
As in the three-qubit case, the eigenvalue problem in Eq.~(\ref{eq:Wstatecondition}) admits multiple half-period solutions including symmetric and antisymmetric Hamiltonians, which are derived in detail in~\ref{app_theory}.
Here, we focus on antisymmetric operations, as they exhibit the shortest synthesis time for a fixed maximal coupling strength.

We consider a chain of $M$ qubits, labelled $\text{q}_m$ for $m=0,\ldots, M-1$.
Because the system's single-excitation Hamiltonian $\hat{H}_M$ has a tridiagonal form, it can be uniquely determined by its eigenvalues $\lambda_m$ and the first component of each eigenstate $|\langle \text{q}_{0}|\lambda_m\rangle|^2$, referred to as the spectral weights of $\hat{H}_M$ ~\cite{vinet2012_PST}.
To ensure periodic dynamics with minimal spectral width, we require the eigenvalues of $\hat{H}_M$ to be linearly spaced, i.e.
\begin{equation}
    \lambda_{m+1} -\lambda_{m}= \frac{\pi}{\tau}, \quad \forall m =0,\ldots, M-2\:.
    \label{eq:generalised_class_spectrum}
\end{equation}
Importantly, this spectrum is characteristic of the Krawtchouk Hamiltonian family, known for realising optimal state transfer in spin-chain systems~\cite{Yung2006_speedPST, Kay2010_PST}.
Although this family provides approximate solutions to the inverse eigenvalue problem, exact W~state generation requires exploring a larger set of Hamiltonians.
We therefore introduce a multi-parameter generalisation of the Krawtchouk Hamiltonians, specified by the eigenvalues~in~Eq.~(\ref{eq:generalised_class_spectrum}) and the spectral weights
\begin{equation}
     |\langle \text{q}_0|\lambda_m\rangle|^2 \propto \frac{1}{m!(M-1-m)!}\prod_{k=1}^{m}\frac{p_k}{1-p_k},
     \label{eq:generalised_class_weights}
\end{equation}
using the convention $\prod_{k=1}^{0}(\cdot) = 1$.
This form contains $M-1$ independent parameters $p_k\in(0,1)$, which uniquely determine $\hat{H}_M$.
Note that choosing \mbox{$p_k=\tilde{p}$} for all $k$ reduces Eq.~(\ref{eq:generalised_class_weights}) to a binomial distribution with probability $\tilde{p}$, corresponding to the spectral weights of the Krawtchouk Hamiltonians~\cite{vinet2012_PST}.
Starting from Krawtchouk Hamiltonians yielding a large overlap with a W state at the half-period $\tau$, we perform numerical searches over the full parameter space $(p_k)$.
This optimisation step yields a finite number of exact solutions, from which we select the instance achieving the shortest synthesis time for a fixed maximal coupling strength (see~\hyperref[sec:met2]{Methods section}).

We experimentally realise the constructed operations for chains with up to $M=7$ qubits.
For the longest chain, an excitation initialised on the central qubit $\text{q}_{6}$ spreads symmetrically to both sides to prepare a seven-qubit W~state within \mbox{$\tau=\SI{264}{\ns}$}, as shown in Fig.~\ref{fig:scaling}\textbf{a}.
The targeted Hamiltonian parameters are provided alongside their fitted experimental values in Fig.~\ref{fig:scaling}\textbf{b}.
Couplings are symmetric in magnitude about the chain centre and lie in the range $\SIrange{1.18}{1.43}{\mega\hertz}$, while on-site energies follow an antisymmetric profile.
In particular, this symmetry is inherited from the standard Krawtchouk Hamiltonians, while our generalisation in Eq.~(\ref{eq:generalised_class_weights}) provides the additional freedom in coupling and energy patterns needed for exact W~state generation.
Additional population dynamics are provided in~\ref{app_all_dynamics}, including different chain lengths and other initialised qubits.

Using QST, the reconstructed seven-qubit state yields an average W state fidelity of \mbox{$F_{\text{W}}=\SI{79.6 \pm 1.3}{\%}$}.
The corresponding results for all implemented operations, including 2D, are summarised in Fig.~\ref{fig:scaling}\textbf{c}.
Beyond the fidelity estimates, each reconstructed state is used to evaluate a witness operator that confirms genuine multipartite entanglement across all cases \cite{haffner2005_ionscalable, Guhne2009_entanglement}, as detailed in~\ref{app_witnesses}.
Crucially, the 2D realisations outperform their 1D counterparts at the same qubit number, despite the greater calibration complexity with more simultaneous couplings.
This improvement stems from the substantially shorter synthesis times achieved in experiment, as indicated by the red and blue markers in Fig.~\ref{fig:scaling}\textbf{d}, thereby reducing the impact of decoherence.
The shaded areas show the expected scaling when the shorter dimension is fixed to \mbox{$L=1$} or \mbox{$L=2$}, with the lower (upper) bounds set by the maximum (minimum) coupling strength measured across all couplers. 
The synthesis time grows linearly with the distance from the initialised qubit to the outermost row or column.
Since the excitation is initialised near the lattice centre, this leads to comparable durations between consecutive even and odd sizes.
Notably, the reported $3\times2$ implementation achieves faster six-qubit entanglement than any 1D operation feasible in our device, underscoring the advantage of single-step protocols that leverage the full available connectivity.

\section*{Discussion}
\label{sec:dis}
We have demonstrated the single-step generation of large W~states via controlled nearest-neighbour interactions in 1D and 2D qubit lattices.
This is achieved by solving the underlying inverse eigenvalue problem to obtain time-independent Hamiltonians that rapidly distribute a single excitation across an entire register.
Using a parametrically-driven superconducting device, we applied this method to directly entangle a $3\times 2$ lattice within $\tau=\SI{99}{\ns}$, achieving a six-qubit W~state fidelity of $\SI{83.9\pm 1.0}{\%}$.
We validated the scalability of the protocol by generating genuine multipartite entanglement in 1D chains of up to seven qubits, with the largest W~state yielding a tomographic fidelity of $\SI{79.6\pm 1.3}{\%}$ for a synthesis time of $\tau=\SI{264}{\ns}$.
In scaled-up systems, these operations can be combined to entangle substantially larger 2D lattices without additional time overhead, since the excitation propagates across both spatial directions in parallel.  

In our implementation, the dominant error sources consist of qubit dephasing during the dynamics, as well as unmitigated SPAM errors affecting QST.
These can be improved in a number of ways.
First, the effect of dephasing can be overcome by increasing the static capacitive couplings $g_{ij/ji}$ between qubits and couplers, thereby achieving stronger parametric interactions and faster dynamics in accordance with Eq.~(\ref{eq:bessel}).
In addition, the pre-tomographic gates used in QST are affected by static ZZ~interactions caused by the hybridisation of higher-excited qubit states.
This effect is particularly strong for specific neighbouring pairs with detunings close to the qubit anharmonicity, \mbox{$|\omega_{\text{q}_i}-\omega_{\text{q}_j}|\approx -\alpha_{\text{q}}\approx\SI{200}{\MHz}$}, where ZZ~couplings reach values on the order of~$\SI{2}{\MHz}$.
These interactions can be mitigated by additional drives~\cite{Noguchi2020_secondorder, Ganzhorn2019_crosstalk, Ni2022_ZZeliminate, Wei2022_crosstalk} or by more precise targeting of the straddling regime \mbox{$|\omega_{\text{q}_i}-\omega_{\text{q}_j}|<-\alpha_{\text{q}}$}, where ZZ~couplings are cancelled at a suitable flux bias~\cite{Mundada2019_ZZsuppression, Sete2021_floatingcoupler, Sung2021_ZZfree}.
Beyond hardware and operational improvements, the measurement overhead associated with QST can be dramatically reduced by using alternative tomographic schemes, such as evaluating a tailored set of $\mathcal{O}(N)$ local observables~\cite{Guehne2007_efficientDetection}, or through parallel estimation of reduced states requiring only $\mathcal{O}(\log(N))$ sequences~\cite{hu2024_effQST}.
These protocols can support scalable fidelity estimation and verification of entanglement in larger systems. 

As both system size and connectivity increase, our protocol offers a fundamental speed advantage over sequential, non-overlapping two-qubit gates. 
Indeed, entangling $N$ qubits in a $d$-dimensional hypercubic lattice demands $\mathcal{O}(N^{1/d}+~\ldots~+N^{1/d})=\mathcal{O}(dN^{1/d})$ layers of two-qubit gates, according to the minimal number of steps for reaching the most distant qubit when moving along the lattice directions.  
By contrast, single-step operations distribute entanglement along all spatial directions in parallel, requiring time $\mathcal{O}(\max(N^{1/d},\ldots, N^{1/d}))=\mathcal{O}(N^{1/d})$, consistent in scaling with the Lieb--Robinson bounds for entanglement distribution~\cite{Bravyi2006_liebrobinson}.
This $\mathcal{O}(1/d)$ improvement guarantees a more pronounced speedup when moving to higher dimensions at fixed $N$ (see \ref{app_speed} for more details).

Beyond the present realisation, our approach can be extended to different types of entanglement and system connectivities.
First, single-step operations can be used to prepare W~states shared among fewer distant qubits, while leaving intermediate qubits depopulated~\cite{Kay2017_generatestates, wang2025_zigzag}.
Indeed, a related Hamiltonian can be constructed within our framework by selecting alternative parameter values in Eq.~(\ref{eq:generalised_class_weights}), targeting a specified initial and final state.
Because a single system may host many such operations, long-range entanglement can be generated directly across a programmable selection of qubits, determined by the applied unitary and initialised node.
Moreover, our framework readily extends to connectivity graphs beyond hypercubic lattices, including treelike and heavy-hex lattices.
In these settings, Hamiltonians for W~state generation can be designed by combining 1D solutions and beam-splitter operations, as described in~\ref{app_heavyhex}.
Finally, a prospective direction is to directly generate entangled states with multiple excitations.
1D~systems provide an accessible setting for this extension, since their higher-excitation dynamics can be mapped to non-interacting fermions via a Jordan--Wigner transformation~\cite{Naegele2022_FST, RoyRomeiro2025_GHZ}.
Such schemes may enable the efficient preparation of other relevant types of entanglement, including Dicke states exhibiting superradiance~\cite{dicke1954_coherence, Wang2020_radiant}.

\section*{Methods}
\subsection*{Effective Hamiltonian in the rotating frame}
\label{sec:met1}
Starting from the driven Hamiltonian in Eq.~(\ref{eq:SW}), we move to the rotating frame defined by the unitary
\begin{equation*}
\hat{U}(t)=
\exp\left[
-it\sum_{i\in V}
(\tilde{\omega}_{\text{q}_i}-\Delta_i)
\ket{\text{q}_i}\bra{\text{q}_i}
\right]\!.
\end{equation*}
The resulting Hamiltonian has diagonal terms \mbox{$ \sum_{i\in V}\Delta_i\ket{\text{q}_i}\bra{\text{q}_i}$}, while hopping operators transform as
\begin{equation*}
\hat{U}^\dagger
|\text{q}_j\rangle\bra{\text{q}_i}
\hat{U}=e^{i[
(\tilde{\omega}_{\text{q}_j}-\Delta_j)-(\tilde{\omega}_{\text{q}_i}-\Delta_i)]t}
|\text{q}_j\rangle\bra{\text{q}_i}.
\end{equation*}
Hence, each near-resonant term
$J_{ij}e^{-i\omega_{\phi_{ij}}t}|\text{q}_j\rangle\bra{\text{q}_i}$ becomes time-independent in the new frame if 
\begin{equation}
    (\tilde{\omega}_{\text{q}_j}-\Delta_j)-(\tilde{\omega}_{\text{q}_i}-\Delta_i)-\omega_{\phi_{ij}}=0,
\label{eqmet:stationary_constraint}
\end{equation}
which fixes the energy offset between this pair of qubits, \mbox{$\Delta_i-\Delta_j=\omega_{\phi_{ij}}-(\tilde{\omega}_{\text{q}_j}-\tilde{\omega}_{\text{q}_i})$}.
Crucially, these energy offsets must be path-consistent so that a globally defined set of $\Delta_i$ obeys Eq.~(\ref{eqmet:stationary_constraint}) for all interacting pairs.
This requires the drive frequencies to satisfy
\begin{equation}
\sum_{\{i,j\}\in L}
\left[\omega_{\phi_{ij}}-(\tilde{\omega}_{\text{q}_j}-\tilde{\omega}_{\text{q}_i})\right]=0,
\end{equation}
for an ordered traversal of any closed-loop $L\subseteq E$ formed by the active couplers. 
Under this condition, and after applying a rotating-wave approximation to remove the remaining fast-oscillating terms, Eq.~(\ref{eq:SW}) reduces to the time-independent Hamiltonian described in Eq.~(\ref{eq:eff}).

\subsection*{Numerical search of extended 1D solutions}
\label{sec:met2}
For a chain of length $M$, we determine Hamiltonians for exact W~state generation using Eq.~(\ref{eq:generalised_class_spectrum})~and~(\ref{eq:generalised_class_weights}), by optimising the $M-1$ independent parameters $p_k\in(0,1)$.
For odd $M$, when the excitation is initialised at the central qubit, the parameter space can be further reduced by imposing
\begin{equation*}
    p_k = p_{M-k}, \quad \forall k= 1,\ldots, \frac{M-1}{2}\:,
\end{equation*}
which restricts the optimisation to Hamiltonians that preserve population symmetry about the chain centre (see \ref{app_theory}).

\begin{figure}[t]
\centering
\includegraphics{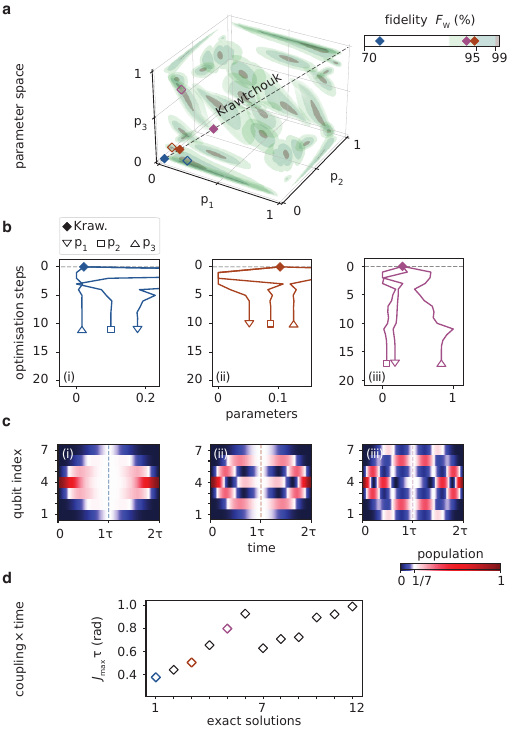}
\caption{
    \label{fig:metnum}
    \textbf{Krawtchouk and exact solutions in a seven-qubit chain.} 
    \textbf{a} Parameter space for symmetry-preserving Hamiltonians with a fixed linear spectrum. Isosurfaces show the W~state fidelity obtained after one half-period $\tau$, for an excitation initialised in the central qubit. The standard Krawtchouk family (dashed line) is highlighted alongside three local optima (solid diamonds), which achieve fidelities in the range $\qtyrange{73.3}{94.4}{\%}$. 
    Each local optimum is located near a distinct solution yielding exact W state generation (hollow diamonds).
    \textbf{b} Numerical optimisation initialised at the highlighted Krawtchouk optima (solid diamonds), revealing the values of $p_1$, $p_2$ and $p_3$ for the different exact solutions (hollow~points). 
    \textbf{c} Population dynamics of all qubits for the obtained operations.
    In all cases, the excitation is delocalised at the half-period $\tau$ with a residual infidelity of $1-F_\text{W}<10^{-9}$. 
    \textbf{d}~Dimensionless synthesis time $J_{\text{max}}\tau$ across all exact operations, where smaller values indicate faster entanglement for a fixed maximal coupling $J_{\text{max}}$. 
    Coloured diamonds indicate the previously highlighted instances.
    The first six solutions are obtained as described above, while the remaining six are found by initialising the optimisation outside the standard Krawtchouk family.
    }
\end{figure}

We consider the $M=7$ case, with the corresponding three-dimensional parameter space shown in Fig.~\ref{fig:metnum}\textbf{a}.
The colour scale indicates the W state fidelity calculated at the half-period, while the diagonal line corresponds to the Krawtchouk family, satisfying \mbox{$p_k=\tilde{p}\in(0,1)$ for all $k$}.
In this subset, the couplings $\tilde{J}_m$ and on-site energies $\tilde{\Delta}_m$ take the explicit analytical form
\begin{equation*}
    \begin{cases}
        \big|\tilde{J}_m(\tilde{p})\big| = \frac{\pi}{\tau}\sqrt{\tilde{p}(1-\tilde{p})m(M-m)} \\
        \tilde{\Delta}_m(\tilde{p}) = \frac{\pi}{\tau}(1-2\tilde{p})\Big(m-\frac{M+1}{2}\Big)\:,
    \end{cases}
    \label{eq:std_kraw}
\end{equation*}
for $m=1,\ldots,M$ \cite{nikiforov1991_cDOP}.
In particular, choosing either $\tilde{p}$ or $1-\tilde{p}$ yields Hamiltonians that are identical up to reflection about the chain centre, since \mbox{$\tilde{J}_{m}(1-\tilde{p})=\tilde{J}_{M-m}(\tilde{p})$} and \mbox{$\tilde{\Delta}_{m}(1-\tilde{p})=\tilde{\Delta}_{M+1-m}(\tilde{p})$}.
Importantly, an initial scan confined to the Krawtchouk family reveals six unique local optima, each achieving a W~state fidelity in the range~$\SIrange{73.3}{94.4}{\%}$.

Next, we turn to the full parameter space, where each Krawtchouk optimum provides an initial point close to a distinct exact solution.
These are then located via numerical optimisation using the Sequential Least Squares Quadratic Programming (SLSQP) algorithm~\cite{Virtanen2020_scipy}, as illustrated in Fig.~\ref{fig:metnum}\textbf{b}.
While all instances prepare a W~state with residual infidelity \mbox{$1-F_{\text{W}}<10^{-9}$}, they exhibit qualitatively distinct dynamics over a period, as summarised in Fig.~\ref{fig:metnum}\textbf{c}.
Correspondingly, the maximal coupling $J_{\text{max}}$ required for a fixed synthesis time~$\tau$ varies substantially, as shown in Fig.~\ref{fig:metnum}\textbf{d}.
Generally, efficient solutions with lower~$J_{\text{max}}\tau$ exhibit a more gradual spreading of the excitation, with fewer ripples in the population dynamics.
We repeat this numerical procedure for all considered chain lengths and select the lowest~$J_{\text{max}}\tau$ operation for hardware implementation.

\section*{Data Availability}
All relevant data supporting the main conclusions and figures of the document are available on request. 

\bibliography{refs}

\newpage
\section*{Acknowledgments}
We thank Yvonne Gao, Hoi-Kwong Lo and Barbara Kraus for insightful discussions.
This work received financial support from the European Union’s Horizon 2020 research, 
the innovation program `MOlecular Quantum Simulations' (MOQS; Nr.~955479),
the EU MSCA Cofund `International, interdisciplinary and intersectoral doctoral program in Quantum Science and Technologies' (QUSTEC; Nr.~847471),
the BMBF programs `German Quantum Computer based on Superconducting Qubits' (GeQCoS; Nr.~13N15680) and MUNIQC-SC (Nr.~13N16188),
the German Research Foundation project `Multi-qubit gates for the efficient exploration of Hilbert space with superconducting qubit systems' (Nr.~445948657) and the excellence initiative `Munich Center for Quantum Science and Technology' (MCQST; Nr.~390814868) as well as the Munich Quantum Valley, which is supported by the Bavarian state government with funds from the Hightech Agenda Bayern Plus. 

\raggedbottom
\section*{Author contributions}
J.R. designed and carried out the experiments and analysed the data.
J.R. and F.R. developed the theoretical framework and performed numerical simulation.
N.B., J.F., L.K. and L.S. fabricated the device.
I.T. and G.H. designed the device.
N.G., M.S., M.W., J.R. and F.R. developed the measurement software framework.
J.S. and F.W. designed and procured the cryoperm shields, the PCB and the device housing.
M.W., C.S., N.G.,  S.S., J.R. and G.H. built and maintained the experimental setup.
S.F. and M.W. supervised the project.

\pagebreak
\onecolumn
\newcommand\myminus{\kern1.5pt\rule[3pt]{4pt}{0.4pt}\kern1pt}
\newcommand\myminustable{\kern1pt\rule[2.2pt]{3pt}{0.3pt}\kern0.5pt}
\sisetup{separate-uncertainty=true,range-units=single,range-phrase=\myminustable}
\renewcommand{\figurename}{Supplementary Fig.}
\renewcommand{\theHfigure}{Supplementary Fig.}
\setcounter{figure}{0}
\renewcommand{\tablename}{Supplementary Table}
\renewcommand{\thesubsection}{Supplementary Note \arabic{subsection}}
\renewcommand{\theHsubsection}{\arabic{subsection}}

\newpage
\section*{Supplementary Notes}
\subsection{-- Experimental setup}
\label{app_setup}
The superconducting device is mounted at the base plate ($\SI{10}{m\kelvin}$) of a dilution refrigerator, as shown in Supplementary~Fig.~\ref{figapp:setup}\textbf{a}.
Control signals are delivered to a routing Printed Circuit Board (PCB) via two 24-channel Ardent multicoaxial connectors.
The PCB is then connected to the 48 device ports via wire bonds, while on-device routing is implemented using air bridges~\cite{bruckmoser2026_nbairbridge}.

\begin{figure*}[b!]
\centering
\includegraphics{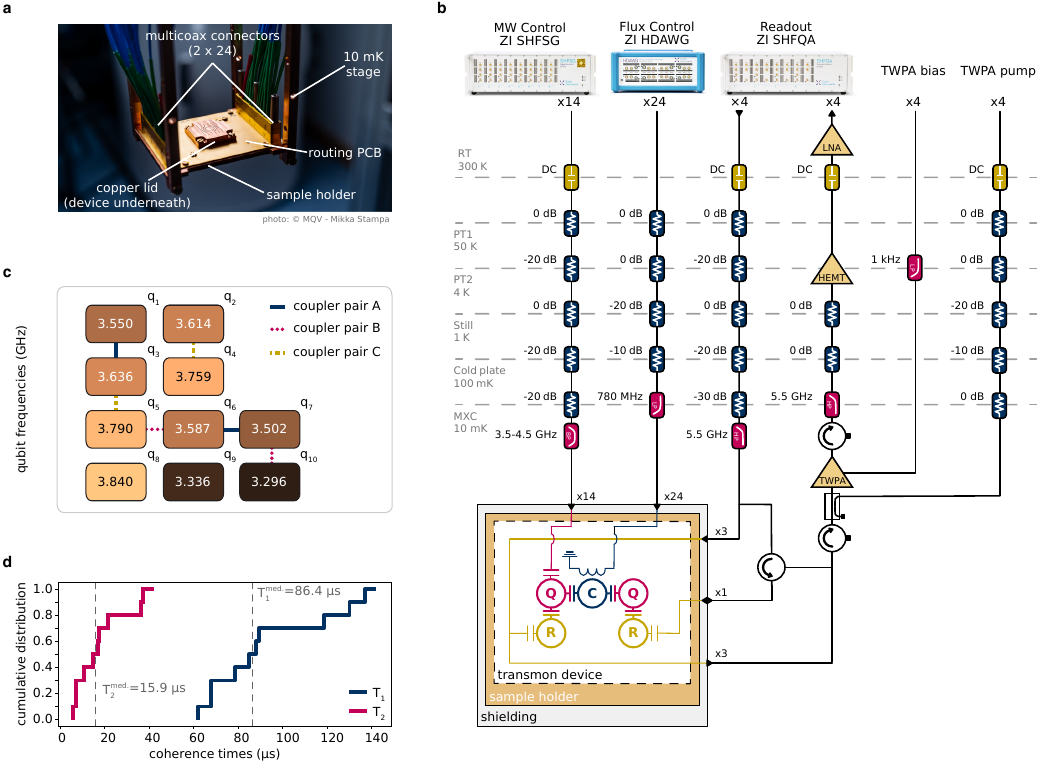}
\caption{
    \label{figapp:setup}
\textbf{Experimental setup, qubit frequencies and coherence times}.
\textbf{a} Photo of the device installed in the base plate of a dilution refrigerator, including copper sample holder, routing PCB and multicoaxial connectors.
\textbf{b} Schematic of experimental setup, showing control components and wiring 
\textbf{c} Transition frequencies of qubits $\text{q}_1$ through $\text{q}_{10}$.
Additional lines mark the three coupler pairs whose drive frequencies differ by less than $\SI{10}{\MHz}$.
\textbf{d} Cumulative distribution of the relaxation times $T_1$ (blue) and echo dephasing times $T_2$ (magenta) measured across qubits $\text{q}_1$ through $\text{q}_{10}$.
Median values are, respectively, $T^{\text{med}}_1\approx\SI{86.4}{\mu s}$ and $T_2^{\text{med}}\approx\SI{15.9}{\mu s}$ (vertical~dashed~lines).
}
\end{figure*}

A schematic of the control setup is shown in Supplementary~Fig.~\ref{figapp:setup}\textbf{b}.
Microwave pulses for single-qubit operations are generated using ``Super High Frequency Signal Generators" (SHFSGs) from Zurich Instruments (ZI), with separate physical channels attributed to each operated qubit.
All microwave control lines are attenuated by a total of \SI{-60}{\dB} distributed across the cryostat stages.
At $\SI{10}{m\kelvin}$, each microwave line is routed through a band-pass filter in the qubit frequency range (Mini-Circuits VBFZ-4000-S+).
Furthermore, flux biasing and AC parametric drives are applied jointly using ZI ``High Density Arbitrary Waveform Generators" (HDAWGs).
Flux lines include a total attenuation of \SI{-30}{\dB} and low-pass filtering with a cutoff frequency of \SI{780}{\mega\hertz} (Mini-Circuits VLFX-780+).
Finally, readout signal generation and output analysis are performed with a ZI ``Super High Frequency Quantum Analyzer" (SHFQA), which enables frequency-multiplexed readout of multiple qubits.
Generated readout signals are attenuated by a total of \SI{-80}{\dB} and high-pass filtered with \SI{5.5}{\GHz} cutoff frequency (Mini-Circuits VHF5050+), thus selecting the resonator band while suppressing noise at qubit frequencies.

In our device, three of the readout feedlines are transmission-based, with distinct input and output ports, while the fourth single-port line is designed for reflection measurements.
The latter configuration requires an additional circulator at MXC, for proper input and output routing. 
The signal output from each feedline is amplified by a dedicated Travelling Wave Parametric Amplifier (TWPA), which is mounted at $\SI{10}{m\kelvin}$ between two isolators.
The second amplification stage is located at $\SI{4}{\kelvin}$ and consists of \SI{40}{\dB} HEMT cryogenic low-noise amplifiers (LNF~LNC4\_8G). Finally, after an additional \SI{45}{\dB} amplification at room temperature (Qotana~DBLNA104000800), the measurement signal is routed back to the SHFQA for demodulation, sampling and analysis.

We now summarise the operating frequencies and coherence times relevant to the protocol implemented in this work.
Experimental qubit frequencies at the operating bias point are shown in Supplementary~Fig.~\ref{figapp:setup}\textbf{c}.
Importantly, the detunings between neighbouring qubit pairs determine the frequencies of the applied parametric drives, as described in the main text.
Spectral crowding among these parametric transitions, in combination with AC flux crosstalk, could lead to leakage outside the target qubit states.
In our device, however, near-degenerate drive frequencies occur only for three pairs of spatially separated couplers, thus suppressing the impact of flux crosstalk. 
Finally, the relaxation~($T_1$) and echo dephasing~($T_2$) times measured across all ten used qubits are shown in Supplementary~Fig.~\ref{figapp:setup}\textbf{d}. 

\subsection{-- Detection of genuine multipartite entanglement}
\label{app_witnesses}

\begin{figure}[b]
\centering
\includegraphics{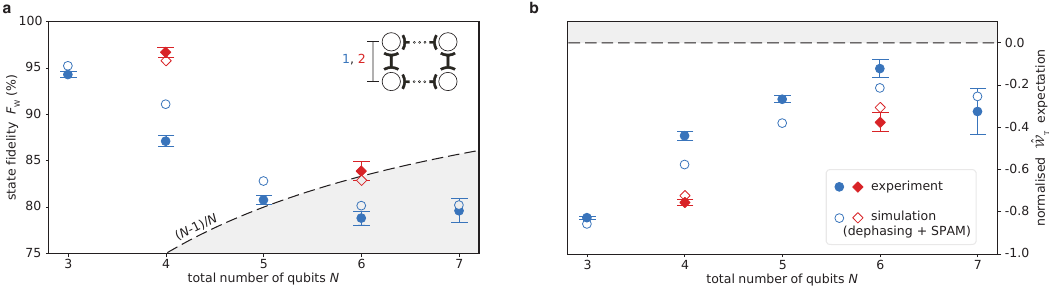}
\caption{
    \label{fig:methods}
    \textbf{Multipartite entanglement detection in different operation dimensions.}
    \textbf{a} W~state fidelity of realised operations with $1\times N$ (blue circles) and $2\times (N/2)$ qubits (red diamonds).
    Solid markers correspond to the experimental tomographic fidelities, while empty markers show simulation results including estimated dephasing and SPAM errors. 
    $N$-qubit entanglement is confirmed for every instance exceeding the $(N-1)/N$ threshold (dashed line).
    \textbf{b} Normalised expectation values of the $\hat{\mathcal{W}}_T$ witness (see text for definition). Negative expectations confirm genuine $N$-qubit entanglement in all instances. 
    }
\end{figure}

To certify that all executed operations generate genuine multipartite entanglement, we evaluate appropriate witness operators on the reconstructed states $\rho_{\text{exp}}$.
An observable $\hat{\mathcal W}$ is an entanglement witness if its expectation value satisfies
\begin{equation}
    \Tr(\hat{\mathcal W}\rho_{\mathrm{sep}})\geq0
    \label{eq:witness_condition}
\end{equation}
for all biseparable states $\rho_{\mathrm{sep}}$, so that $\Tr(\hat{\mathcal W}\rho_{\text{exp}})<0$ confirms multipartite entanglement~\cite{Guehne2009_entanglementDetection, chruscinski2014_witnesses}.

For a W~state with $N$ qubits, the maximal overlap achieved with any biseparable state is $(N-1)/N$~\cite{Guehne2007_efficientDetection}.
Hence, an operator testing whether the state fidelity $F_{\text{W}}=\bra{W_N}\rho_{\text{exp}}\ket{W_N}$ exceeds this threshold,
\begin{equation}
    \hat{\mathcal{W}}_F = \frac{N-1}{N}\hat{\mathbb{I}}_N-\ket{W_N}\bra{W_N}\!, 
    \label{eq:witness_f}
\end{equation}
is a valid witness~\cite{Guehne2007_efficientDetection}. 
Here, $\hat{\mathbb{I}}_N$ denotes the identity on the $N$-qubit Hilbert space.
Applied to our experimental dataset, $\hat{\mathcal W}_F$ is sufficient to confirm genuine entanglement of up to five qubits in 1D and six qubits in 2D, as shown in Supplementary~Fig.~\ref{fig:methods}\textbf{a}.

For the largest 1D implementations with $N=6$ and $7$, a more selective entanglement witness is required to suppress dominant noise channels and exclude specific biseparable states.
One such operator form, proposed in~\cite{haffner2005_ionscalable}, is
\begin{equation}
    \hat{\mathcal{W}}_T(\beta) =  \gamma(\beta) \hat{\mathbb{I}}_N 
    - \ket{\text{W}_N}\bra{\text{W}_N} 
     + \beta \sum_{n=1}^N 
    \ket{0_n}\bra{0_n} \otimes 
    \ket{\text{W}_{N-1}}\bra{\text{W}_{N-1}}\!, 
   \label{eq:witness_T}
\end{equation}
where the additional term $\propto\beta$ accounts for biseparable states of the form $\ket{0_n}\otimes\ket{\text{W}_{N-1}}$, with qubit $n$ in the ground state and the remaining qubits sharing a W~state.
For every experimental state $\rho_{\text{exp}}$, we select $\beta\in\mathbb{R}$ to minimise the expectation value $\Tr(\hat{\mathcal{W}}_T(\beta)\rho_{\text{exp}})$, while choosing $\gamma(\beta)\in\mathbb{R}$ accordingly to satisfy Eq.~(\ref{eq:witness_condition}), such that $\hat{\mathcal{W}}_T(\beta)$ is a valid witness~\cite{haffner2005_ionscalable, Guhne2009_entanglement}.  
Finally, $\hat{\mathcal{W}}_T(\beta)$ is renormalised so that the mixed state $\hat{\mathbb{I}}_N / 2^N$ yields unit expectation value, thus enabling direct comparison of all measured values, which are shown in Supplementary~Fig.~\ref{fig:methods}\textbf{b}.
The expectation values are exclusively negative, confirming genuine multipartite entanglement across all cases.

\subsection{-- Inverse eigenvalue method and families of 1D operations}
\label{app_theory}

Here, we derive in detail three distinct families of Hamiltonians enabling single-step W~state generation in 1D chains.
These comprise symmetric and antisymmetric operations, as well as solutions restricted to resonant interactions~\cite{Perezleija2013_multiportw, Vildoso2023_lowlossWGAs, bugarski2024_waveguidew}.
Crucially, their description is based on the standard theory of discrete orthogonal polynomials~\cite{nikiforov1991_cDOP}, which provides a systematic framework for reconstructing a 1D Hamiltonian from prescribed spectral properties~\cite{Kay2017_generatestates, petrovic2018_opticaltransfer, wang2025_inverse}.
Because the spectral structure solving the state generation problem is not unique, different choices ultimately give rise to the distinct families presented here.

Let us consider an $(N+1)$-qubit chain with nearest-neighbour $XX+YY$ couplings.
In the subspace spanned by the localised single-excitation states $\{\ket{\text{q}_n}\}_{0\leq n\leq N}$, the Hamiltonian takes the tridiagonal form 
\begin{equation}
H_{N+1} =
\begin{pmatrix}
\Delta_0 & J_1      & 0        & \cdots   & 0 \\
J_1      & \Delta_1 & J_2      & \ddots   & \vdots \\
0        & J_2      & \Delta_2 & \ddots   & 0 \\
\vdots   & \ddots   & \ddots   & \ddots   & J_N \\
0        & \cdots   & 0        & J_N      & \Delta_N
\end{pmatrix}\!,
\label{eqapp:tridiag}
\end{equation}
where $\Delta_n$ and $J_n$ denote the local energy and coupling at position $n$, respectively.
Importantly, the complex arguments $\arg(J_n)$ do not impact the dynamics (see \ref{app_AB}), and all couplings $J_n$ are chosen real without loss of generality. 
Furthermore, let \mbox{$\{\ket{\lambda_m}\}_{0\leq m\leq N}$ denote the eigenbasis of $H_{N+1}$}, with eigenvalues ordered as \mbox{$\lambda_0 < \lambda_1 <\ldots < \lambda_{N}$}.
Projecting Eq.~(\ref{eqapp:tridiag}) onto each of the eigenstates $\ket{\lambda_m}$ reveals clear recurrence relations between the overlap coefficients $C_{m,n} \coloneqq\langle \lambda_m | \text{q}_n\rangle$,
\begin{equation}
\begin{cases}    
\lambda_mC_{m,n} = J_nC_{m,n-1} + \Delta_nC_{m,n} + J_{n+1}C_{m,n+1}\quad \forall m,n=0,1,\ldots,N,
\end{cases}
\label{eqapp:recurrence1}
\end{equation}
where we use the boundary conditions $J_0=J_{N+1}=0$. 
We now introduce a family of real polynomials $\chi_n:\mathbb{R}\to\mathbb{R}$, defined as
\begin{equation}
    \chi_n(\lambda_m) \coloneqq \frac{C_{m,n}}{C_{m,0}}
    \quad \forall m,n=0,1,\ldots,N.
    \label{eqapp:poly_def}
\end{equation}
From the relation $\sum_m C_{m,n}C_{m,i}=\delta_{n,i}$, where $\delta$ denotes the Kronecker delta, it follows that these polynomials are orthogonal with respect to the inner product
\begin{equation*}
(\chi_n, \chi_{i})\coloneqq\sum_{m=0}^{N}w_m\: \chi_n(\lambda_m)\:\chi_{i}(\lambda_m) = \delta_{n,i} \quad \forall n, i \in 0,1,\ldots,N,
\end{equation*}
where the discrete weights $w_m=(C_{m,0})^2$ satisfy the normalisation condition $\sum_m w_m = 1$.
Moreover, each polynomial $\chi_n$ is of degree $n$, as follows by inserting Eq.~(\ref{eqapp:poly_def}) into the recurrence relation in~Eq.~(\ref{eqapp:recurrence1}),
\begin{equation}   
\lambda \chi_n(\lambda) = J_n\chi_{n-1}(\lambda) + \Delta_n\chi_n (\lambda) + J_{n+1}\chi_{n+1}(\lambda)\quad \forall n=0,1,\ldots,N,
\label{eqapp:recurrence2}
\end{equation}
with the initial conditions $\chi_{-1}=0$ and $\chi_0=1$~\cite{vinet2012_PST}. Lastly, taking the inner product of both sides of Eq.~(\ref{eqapp:recurrence2}) with $\chi_n$ yields
\begin{equation}
\begin{cases}
    \Delta_n = (\chi_n, \:\lambda\chi_n) \\
\end{cases}\quad\forall n=0,1,\ldots,N,
\label{eqapp:parameters_product_1}
\end{equation}
while the inner product with $\chi_{n-1}$ yields
\begin{equation}
\begin{cases}
    J_n = (\chi_{n-1}, \:\lambda\chi_n)
\end{cases}\quad\forall n=1,\ldots,N,
\label{eqapp:parameters_product_2}
\end{equation}
Eq.~(\ref{eqapp:recurrence2}),~(\ref{eqapp:parameters_product_1})~and~(\ref{eqapp:parameters_product_2}) are central to reconstructing $H_{N+1}$, as they suffice to compute all polynomials $\chi_n$ and Hamiltonian parameters $\Delta_n$ and $J_n$, once provided the eigenvalues $\lambda_m$ and weights $w_m$~\cite{wang2025_inverse}. 
As such, different families of solutions to our entanglement problem are defined by distinct parametrisation choices of $\lambda_m$ and $w_m$.

\begin{figure*}[t!]
\centering
\includegraphics{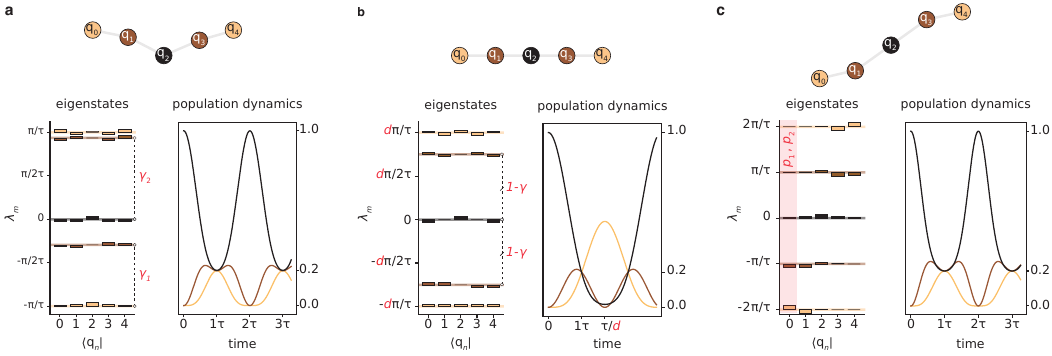}
\caption{
    \label{figapp:poly}
\textbf{Symmetric, resonant and antisymmetric Hamiltonians for exact W~state generation in 1D}.
Each~panel shows a representative five-qubit chain from one Hamiltonian family.
The top row depicts the on-site energy profile along the chain.
The bar plots represent the components $\bra{\text{q}_n}\lambda_m\rangle$ of the eigenstates $\ket{\lambda_m}$, which are offset by the corresponding eigenvalues $\lambda_m$.
The bottom-right plots show the population dynamics when the central qubit, $\text{q}_2$, is initialised.
Elements highlighted in red indicate the free parameters that span each Hamiltonian family (see text for details).
\textbf{a} Hamiltonians in the symmetric family are designed by fixing every second eigenvalue to a linear grid, while the remaining eigenvalues are placed with relative gaps $\gamma_1$ and $\gamma_2$.
\textbf{b} Resonant Hamiltonians additionally exhibit a spectrum that is reflection-symmetric about $0$, such that the relative gaps follow $\gamma_2=1-\gamma_1$.
\textbf{c} Antisymmetric Hamiltonians are constructed with strictly linear spectra, while the corresponding spectral weights (defined as the population on the first qubit for each eigenvector) vary according to the free parameters $p_1$ and $p_2$.
}
\end{figure*}

The first case we consider 
consists of mirror-symmetric Hamiltonians, i.e. respecting $J_n =J_{N-n+1}$ and \mbox{$\Delta_n = \Delta_{N-n}$} for all~$n$.
Under this symmetry constraint, the weights are uniquely determined by the eigenvalues~\cite{vinet2012_PST}, according to
\begin{equation}
w^{\text{sym}}_m
\propto
\left.\frac{(-1)^{N+m}}
{
\displaystyle
\frac{\partial}{\partial \lambda}
\prod_{i=0}^{N} (\lambda - \lambda_i)
}\right|_{\lambda=\lambda_m}.
\label{eqapp:symmetric_weights}
\end{equation}
As an important consequence, the eigenvalues alone fully specify any mirror-symmetric Hamiltonian.
Moreover, the corresponding eigenstates $\ket{\lambda_m^{\text{sym}}}$ must follow the property
\begin{equation*}
\begin{cases}
\bra{\text{q}_n}\lambda_m^{\text{sym}}\rangle = (-1)^{m} \bra{\text{q}_{N-n}}\lambda_m^{\text{sym}}\rangle\quad\forall m,n=0,1,\ldots,N ,    
\end{cases}
\end{equation*}
thus partitioning the eigenbasis into alternating symmetric (even-indexed) and antisymmetric (odd-indexed) states, with respect to the centre of the chain~\cite{Kay2010_PST}.
For a symmetric initial state $\ket{\psi_{0}}$, such as a centralised excitation in an odd-length chain, the dynamics are confined to the subspace spanned by $\{\ket{\lambda_{2m}^{\text{sym}}}\}$, since $\langle\lambda_{2m-1}^{\text{sym}}\ket{\psi_{0}}=0$ for all $m$.
We therefore define a family of mirror-symmetric Hamiltonians on a chain with odd length, each uniquely specified by the spectrum
\begin{equation}
\begin{cases}
\lambda^{\text{sym}}_{2m} =\dfrac{\pi}{\tau}\left(m-\dfrac{N}{2}\right)
& m=0,\ldots,\dfrac{N}{2},\\[6pt]
\lambda^{\text{sym}}_{2m-1} = \gamma_m\,\lambda^{\text{sym}}_{2m}+(1-\gamma_m)\,\lambda^{\text{sym}}_{2m-2}
& m=1,\ldots, \dfrac{N}{2}.
\end{cases}
\label{eqapp:symmetric_spectrum}
\end{equation}
where the even-indexed eigenvalues $\lambda^{\text{sym}}_{2m}$ are placed in a linear grid, while the parameters $\gamma_m\in (0,1)$ specify the relative locations of the odd-indexed eigenvalues $\lambda^{\text{sym}}_{2m-1}$, without violating the ordering.
This spectrum choice guarantees $2\tau$-periodic dynamics for any symmetric initial state, while the free parameters $\gamma_m$ are used to shape the dynamics over each period. 
Supplementary~Fig.~\ref{figapp:poly}\textbf{a} shows a five-qubit Hamiltonian constructed with Eq.~(\ref{eqapp:symmetric_spectrum}) for exact W~state generation at $\tau$, where $\gamma_1\approx70.5411\times10^{-2}$ and $\gamma_2\approx93.5872\times10^{-2}$.
Notably, Eq.~(\ref{eqapp:symmetric_spectrum}) constitutes a multi-parameter generalisation to the Para-Krawtchouk Hamiltonians, originally proposed for the study of fractional state revival in spin chain systems~\cite{genest2016_fractional}.

The second considered case consists of odd-length chains with purely resonant couplings, i.e. with uniform on-site energies $\Delta_n =0$ for all $n$.
If $J_n=J_{N-n+1}$, the Hamiltonian is again fully determined by its eigenvalues $\lambda_m^{\text{res}}$, with weights obtained from Eq.~(\ref{eqapp:symmetric_weights}).
In this resonant setting, the full spectrum must be reflection-symmetric, i.e.
\begin{equation*}
\lambda^\text{res}_m = -\lambda^\text{res}_{N-m}
\quad\forall m=0,\ldots,\frac{N}{2},
\end{equation*}
up to a trivial global offset.
Combining this condition with Eq.~(\ref{eqapp:symmetric_spectrum}) leads to the parametrisation 
\begin{equation}
\begin{cases}
\lambda^\text{res}_{2m} =\dfrac{d\pi}{\tau}\left(m-\dfrac{N}{4}\right)
& m=0,\ldots,\left\lfloor\dfrac{N}{4}\right\rfloor\!,\\[8pt]
\lambda^\text{res}_{2m-1}
= \gamma_m\,\lambda^\text{res}_{2m}+(1-\gamma_m)\,\lambda^\text{res}_{2m-2}
& m=1,\ldots,\left\lfloor\dfrac{N+2}{4}\right\rfloor\!,\\[8pt]\lambda^\text{res}_{m} = -\lambda^\text{res}_{N-m}
& m=\dfrac{N}{2}+1,\ldots,N,
\end{cases}
\label{eqapp:resonant_spectrum_param}
\end{equation}
where the free parameters $\gamma_m\in(0,1)$ once again determine the relative spectral gaps, and $d\in(0, 1]$ denotes a global scaling factor, as illustrated in Supplementary~Fig.~\ref{figapp:poly}\textbf{b}.
Once again, the example shown is constructed to prepare a five-qubit W~state in time $\tau$, which requires $\gamma_1\approx 24.8504\times10^{-2}$ and $d\approx58.9178\times10^{-2}$.
Despite the different derivation, the operations obtained from Eq.~(\ref{eqapp:resonant_spectrum_param})
are equivalent to those proposed in~\cite{Perezleija2013_multiportw} for 1D optical waveguide arrays. 

Finally, the third considered set of solutions generally exhibits an antisymmetric parameter structure, with \mbox{$J_n =-J_{N-n+1}$} and $\Delta_n = -\Delta_{N-n}$ for all $n$.
Crucially, Eq.~(\ref{eqapp:symmetric_weights}) does not apply in this regime, and the correspondence between eigenvalues $\lambda^{\text{anti}}_{m}$ and weights $w^\text{anti}_m$ is no longer unique.
We specifically consider isospectral Hamiltonians satisfying
\begin{equation}
\begin{cases}
     \lambda^{\text{anti}}_{m} =\frac{\pi}{\tau}\left(m-\frac{N}{2}\right)\\[6pt]
     w^\text{anti}_m \propto \frac{1}{m!(N-m)!}\prod_{k=1}^{m}\frac{p_k}{1-p_k}
\end{cases}\quad\forall m=0,\ldots,N,
     \label{eqapp:antisymmetric_weights}
\end{equation}
using the product convention $\prod_{k=1}^{0}(\cdot) = 1$.
This choice of fully linear spectrum guarantees a period of $2\tau$ for any initial state, while the free parameters $p_k\in(0,1)$ modify the weight distribution, thereby shaping the eigenstates and the dynamics within each period.
For a five-qubit chain, the parameters $p_1=p_4=13.7471\times10^{-2}$ and $p_2~=~p_3~=~2.7229\times10^{-2}$ yield exact W~state generation, as shown in Supplementary~Fig.~\ref{figapp:poly}\textbf{c}.
Unlike the previous symmetric and resonant constructions, Eq.~(\ref{eqapp:antisymmetric_weights}) does not assume a symmetric initial state and can therefore be used to design solutions with off-centred initial states, including the even-length operations demonstrated in this work.
As explored in the main text, Eq.~(\ref{eqapp:antisymmetric_weights}) acts as a multi-parameter extension to a well-known Hamiltonian family, since setting $p_k=\tilde{p}\in(0,1)$ for all $k$ recovers the standard binomial distribution $w_m\sim B(N, \tilde{p})$ associated with the Krawtchouk Hamiltonians~\cite{Yung2006_speedPST, Kay2010_PST}.

\subsection{-- Robustness of single-step operations to parameter variations}
\label{app_robustness}
We numerically study the performance of the different 1D operations derived in \ref{app_theory} under Hamiltonian parameter variations.
More specifically, we introduce additive Gaussian noise to all Hamiltonian parameters affecting the dynamics,
\begin{equation*}
    \Delta_n \rightarrow \Delta_n + \delta \Delta_n,
    \qquad
    |J_n| \rightarrow |J_n| + \delta J_n,
\end{equation*}
where each perturbation $\delta \Delta_n$ and $\delta J_n$ is sampled independently from a normal distribution with standard deviation~$\sigma$.
For a fair comparison, the maximal coupling strength $J_{\text{max}}\coloneqq\max_n \big(|J_n|\big)$ is considered fixed across all operations.

Using $\sigma=0.08J_{\text{max}}$, we simulate the excitation dynamics of different five-qubit Hamiltonians using $S=450$ independent noise samples in each case, with results summarised in Supplementary~Fig.~\ref{figapp:robust}\textbf{a}.
While the mean trajectories approximate the corresponding noiseless dynamics (see Supplementary~Fig.~\ref{figapp:poly}), the spread differs significantly across all cases.
This contrast is captured by the state delocalisation
\begin{equation}
    \mathcal{D}\coloneqq\left(N\sum_{n=1}^{N} P_n^2\right)^{-1}\!,
\end{equation}
where $N=5$ and $P_n$ denotes the population of qubit $n$ measured at the synthesis time.
The symmetric and antisymmetric operations exhibit consistently higher values of $\mathcal{D}$, as the corresponding dynamics have zero time derivative at the synthesis time, indicating a more robust excitation distribution.
This advantage persists over different values of parameter deviation $\sigma$, as shown in Supplementary~Fig.~\ref{figapp:robust}\textbf{b}.

Importantly, $\mathcal{D}$ quantifies the uniformity of the excitation distribution, independently of its phase coherence.
Under slow parameter drifts, the final state remains a largely coherent superposition, equivalent to an ideal W~state up to local phase offsets that are easily corrected with virtual $Z$ rotations.
Therefore, $\mathcal{D}$ provides a suitable measure of robustness in this regime.
By contrast, parameter fluctuations occurring on the timescale of individual experimental shots lead to effective dephasing.
In this regime, the W~state fidelity $F_{\text{W}}\coloneqq\bra{W_N}\rho\ket{W_N}$, where $\rho$ denotes the mixed state averaged over the noise distribution, provides the more complete figure of merit.
In our simulations, $F_{\text{W}}$ reveals a more balanced performance across the three operations, while the antisymmetric case remains the most robust, as shown in Supplementary~Fig.~\ref{figapp:robust}\textbf{c}.

\begin{figure*}[t!]
\centering
\includegraphics{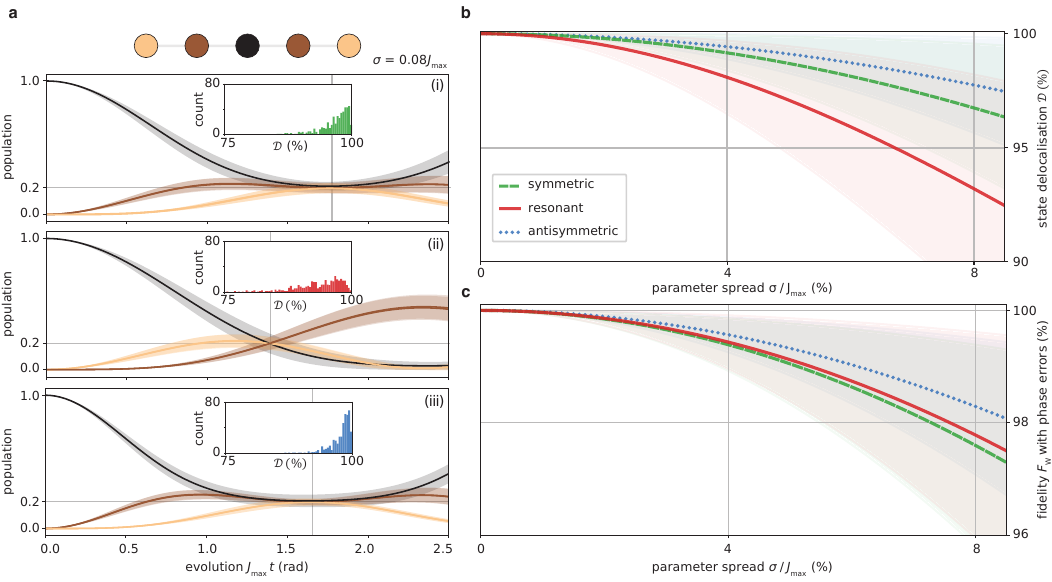}
\caption{
    \label{figapp:robust}
\textbf{Robustness of different single-step operations to parameter variations}. 
\textbf{a} Simulated dynamics of a five-qubit chain undergoing symmetric (i), resonant (ii) and antisymmetric (iii) W~state generation.
Each Hamiltonian parameter is subject to independent additive noise, which is sampled from a normal distribution with standard deviation $\sigma=0.08J_{\text{max}}$, where $J_{\text{max}}$ is the maximal implemented coupling.
Solid lines show the mean over all $S=450$ samples, while shaded areas indicate the standard deviation at each time.
Histograms (insets) show the distribution of the state delocalisation $\mathcal{D}$ calculated at the synthesis time (vertical grey line).
\textbf{b} State delocalisation $\mathcal{D}$ as a function of parameter spread $\sigma$.
Lines and shaded areas respectively indicate the mean and standard deviation, estimated from $450$ samples for each value of $\sigma$.
\textbf{c} W~state fidelity $F_\text{W}$ averaged over all samples, which includes dephasing caused by the parameter fluctuations.
}
\end{figure*}

\subsection{-- Comparison of speed with two-qubit gate decomposition}
\label{app_speed}

Single-step operations provide a speed improvement for W~state generation, when compared with circuits of sequential two-qubit gates.
Here, we show that this advantage becomes more pronounced as the system connectivity increases, since circuit-based schemes are fundamentally limited to at most one active interaction per qubit at a given time.

Let us first consider a 1D chain comprised of $N\geq 3$ qubits.
We prepare an excitation as close as possible to the geometric centre, while all other qubits are initialised in the ground state.
Single-step W~state generation requires time~$\mathcal{O}(\lfloor N/2 \rfloor)$, consistent with the maximal distance from the initialised qubit.  
In contrast, an efficient circuit demands~$\lceil N/2\rceil$ layers of two-qubit gates, following the number of steps required for the excitation to spread to both outermost qubits~\cite{Baertschi2022_shortcircuit}.
In this scenario, each two-qubit gate acts on the corresponding single-excitation manifold $\{\ket{10}, \ket{01}\}$ as
\begin{equation*}
P_{XX+YY}(\theta)=
\begin{pmatrix}
\cos(\frac{\theta}{2}) &  \sin(\frac{\theta}{2}) \\
 -\sin(\frac{\theta}{2}) & \cos(\frac{\theta}{2}) 
\end{pmatrix}\!,
\end{equation*}
where the transfer angle $\theta\in[\pi/4, \pi/2]$ depends on the gate's position in the circuit.
A native implementation of $P_{XX+YY}(\theta)$ can be achieved with optimal $\mathcal{O}(\theta)$ execution, given existing control techniques~\cite{chen2025_efficient2q}.
A speed comparison between single-step operations and the optimal native-gate decomposition is given in Supplementary~Fig.~\ref{figapp:speedup}\textbf{a}, where we assume a global maximal coupling of $J_\text{max} = \SI{2}{\MHz}$.
Excluding the $N=4$ case, the single-step method achieves faster W~state generation, with details of the different reported operations provided in \ref{app_theory}.

When moving to lattices with higher connectivity, an optimal two-qubit gate decomposition is generally unknown, as its determination is analogous to an NP-complete problem: the one-to-all broadcast problem~\cite{jansen1994_broadcastNP}.
Regardless, the minimal number of circuit layers $T_{\text{min}}$ obeys a simple lower bound, set by two fundamental limits.
The first limit is determined by the shortest-path distance from the initialised qubit to the most distant qubit, $T_{\text{min}}\geq d_{1;\max}$.
In a square lattice, $d_{1;\max}$ corresponds to the Manhattan distance, where propagation is restricted to one coordinate at a time, as shown in Supplementary~Fig.~\ref{figapp:speedup}\textbf{b}.
More generally, for an $L\times L$ lattice, one has $d_{1;\max}=2\:\lfloor L/2\rfloor$.
In contrast, the scaling of single-step 2D operations is bounded by the Chebyshev distance $d_{\infty;\max}=\lfloor L/2\rfloor$, where simultaneous propagation on both axes is allowed, thus reaching a larger region in a comparable time frame, as illustrated in Supplementary~Fig.~\ref{figapp:speedup}\textbf{c}.
A second fundamental speed limit imposed on sequential circuits is that
each qubit can spread entanglement to at most one neighbour per step.
As a result, the number of entangled qubits can at most double at each step, implying $T_{\text{min}}\geq \lceil\log_2(N)\rceil$ for a system of $N$ qubits.
Together, these constraints yield the lower bound
\begin{equation}
    T_{\text{min}}\geq \max(\lceil\log_2(N)\rceil, d_{1;\text{max}}).
    \label{eqapp::boundedsteps}
\end{equation}
The corresponding minimum time is shown in Supplementary~Fig.~\ref{figapp:speedup}\textbf{d}, where we assume a time-optimal implementation of the $P_{XX+YY}(\theta)$ gates and a maximal transfer angle of $\theta_{\max}= \pi/2$ in each circuit step.
Single-step operations provide a substantially greater speedup in 2D than in 1D, owing to the more favourable scaling compared to two-qubit gates.

As a final remark, we note that a sequence of smaller $K\times K$ single-step operations, with $K<L$, still reproduces the favourable Chebyshev-distance scaling.
Therefore, this approach remains advantageous even when practical constraints prevent simultaneous control across the entire lattice.
Moreover, single-step methods can be re-optimised for hardware imperfections, including couplings that are unusually weak or entirely absent, as demonstrated in~\cite{Xiang2024_2Dstatetransfer}.

\begin{figure*}[t!]
\centering
\includegraphics{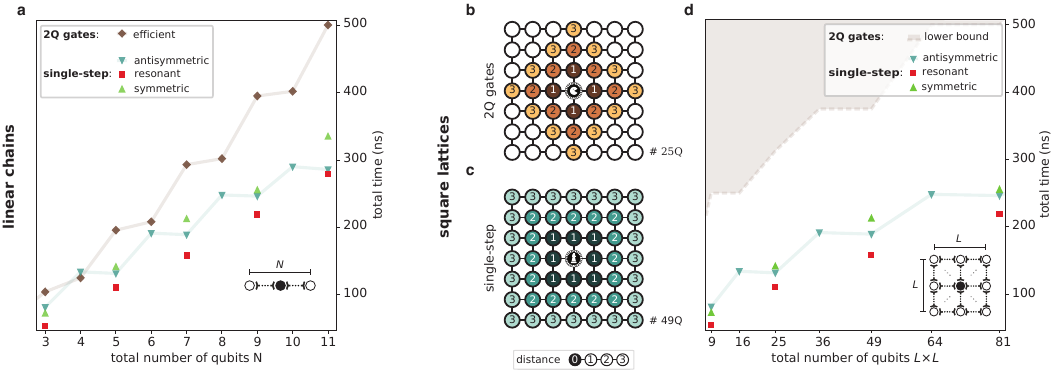}
\caption{
    \label{figapp:speedup}
\textbf{Speed comparison between single-step entanglement and two-qubit gate decomposition}.
\textbf{a} Time scaling of W~state generation in 1D lattices with $N$ qubits, assuming a global maximal coupling of $J_{\text{max}}= \SI{2}{\MHz}$.
Simulation data includes efficient two-qubit gate circuits (brown diamonds)~\cite{chen2025_efficient2q}, as well as antisymmetric (cyan triangles), resonant (red squares) and symmetric (green triangles) single-step operations.
\textbf{b} Entanglement distribution using two-qubit gates in a $7\times 7$ square lattice, with individual qubits shown as circles.
For a fixed number of circuit layers, the accessible region is confined to a diamond shape, set by the Manhattan distance from the initialised qubit.
\textbf{c} Single-step entanglement distribution in the same lattice.
Following the Chebyshev distance, entanglement spreads along the two lattice axes simultaneously, thus covering a larger region within a comparable time.
\textbf{d} Time scaling for W~state generation in 2D square lattices with $N=L\times L$ qubits, where $J_{\text{max}}= \SI{2}{\MHz}$.
The brown area indicates the lower bound for circuits with two-qubit gates (see text for derivation).
Single-step operations (other colours) offer a substantially larger speedup in 2D compared to 1D.
}
\end{figure*}

\subsection{-- Aharonov-Bohm effect in closed loops}
\label{app_AB}
By introducing a phase offset $\varphi_{ij}$ to an AC parametric drive, the corresponding coupling acquires a complex argument, $J_{ij} \rightarrow J_{ij} e^{i\varphi_{ij}}$.
In loop-free connectivities, coupling phases do not affect the excitation dynamics, since they can be removed by redefining the qubit frames as $\ket{\text{q}_i} \rightarrow e^{i\theta_i}\ket{\text{q}_i}$, where $\theta_j-\theta_i=\varphi_{ij}$.
In contrast, the sum of phases around a closed loop is independent of the qubit frame.
This gauge-invariant quantity sets the relative phase between alternative excitation pathways, analogous to the Aharonov--Bohm phase acquired by a charged particle encircling a magnetic flux~\cite{Rosen2024_vectorpotential}. 

We operate a $2\times 2$ section of our device with all couplings set to $|J|\approx2\pi\times\SI{1.5}{MHz}$, as illustrated in Supplementary~Fig.~\ref{figapp:AB}\textbf{a}.
We prepare an excitation on qubit $\text{q}_1$ and record the dynamics for different coupling phases~\mbox{$\varphi\in[-\pi,\pi]$} between qubits $\text{q}_3$ and $\text{q}_4$, with all other phases set to zero. 
The resulting population transfer between $\text{q}_1$ and the diagonally opposite qubit $\text{q}_4$ is shown in Supplementary~Fig.~\ref{figapp:AB}\textbf{b}.
For $\varphi=\pm\pi$, destructive interference between the two excitation trajectories suppresses transfer to $\text{q}_4$, causing the excitation to refocus on $\text{q}_1$.
Conversely, for $\varphi=0$, the interference is strictly constructive, resulting in complete population transfer in $T_\pi=\pi/2|J|=\SI{167}{\ns}$.
This constructive configuration also generates a four-qubit W~state at optimal time $T_\pi/2=\SI{83.5}{\ns}$, as highlighted in Supplementary~Fig.~\ref{figapp:AB}\textbf{c}, here with a tomographic fidelity of $F_{\text{W}}=\SI{96.7\pm 0.5}{\%}$.
More generally, for all entangling 2D operations reported in this work, we tune a sufficient subset of coupling phases so that the sum over each closed loop vanishes, therefore enforcing constructive interference between alternative pathways.

\begin{figure*}[t!]
\centering
\includegraphics{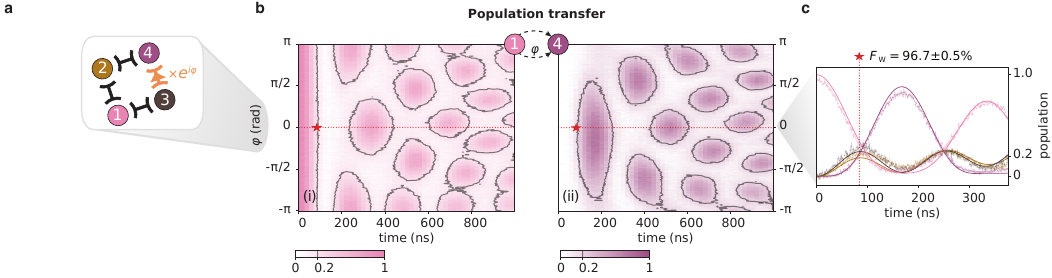}
\caption{
    \label{figapp:AB}
\textbf{Aharonov-Bohm interference and W~state generation in a $2\times 2$ lattice}. \textbf{a} Diagram showing the $2\times 2$ lattice formed by qubits $\text{q}_1$ through $\text{q}_4$, with hopping interactions engineered along all edges. Three parametric drives are applied with fixed phases (black), while the remaining drive phase is swept over the range $\varphi\in [-\pi, \pi]$ (orange). \textbf{b} Population transfer from qubit $\text{q}_1$~(i) to the diagonally opposite qubit $\text{q}_4$~(ii), as a function of the evolution time and $\varphi$. 
The red star marks the operation selected for fast W~state generation ($\varphi=0$).
\textbf{c} Population dynamics recorded for all qubits at $\varphi = 0$. A four-qubit W~state is generated within $\SI{83.5}{\ns}$, with a tomographic fidelity of $F_{\text{W}}=\SI{96.7\pm 0.5}{\%}$.
}
\end{figure*}

\subsection{-- Single-step W~state generation in other regular connectivities}
\label{app_heavyhex}

As demonstrated in the main text, 1D operations can be combined to directly generate entanglement in square lattices. More generally, such 1D Hamiltonians provide building blocks for entangling a broader range of connectivity graphs, particularly when combined with beam-splitter operations~\cite{perales2005manipulating, Yang2006_beamsplitter, Ross2011_spinnetworks}.

Let us consider a graph $G \coloneqq (V,E)$ whose vertices $V$ represent qubits and edges $E$ denote interacting pairs.
The corresponding system is initialised with a single excitation on $\text{q}_0^0 \in V$, while all other qubits are prepared in the ground state. The set of qubits can be partitioned as $V=\bigoplus_{d=0}^D V_d$, where each layer $V_d \coloneqq \{\text{q}_d^k\}_{0 \leq k \leq K_d-1}$ contains all $K_d$ qubits at distance $d$ from $\text{q}_0^0$.
We take $G$ to be a distance-regular graph rooted at $\text{q}_0^0$, i.e. every qubit in a layer $V_d$ has precisely $L_d'$ neighbours within that layer and $L_d$ neighbours in the next layer $V_{d+1}$.
One notable example is the heavy-hex lattice shown in Supplementary~Fig.~\ref{figapp:heavyhex}\textbf{a}, which is rooted at the centremost qubit $\text{q}_0^0$, with $L_d' = 0$ and $L_d \in \{1,2,3\}$ for every $d=0,1,\ldots,6$.

By engineering Hamiltonian parameters that depend only on the distance $d$, the dynamics remain confined to the subspace spanned by the layer superposition states
\begin{equation*}
\ket{\text{W}_d} \coloneqq \frac{1}{\sqrt{K_d}} \sum_{k=0}^{K_d-1} \ket{\text{q}_d^k}\!.
\end{equation*}
In this basis, the Hamiltonian takes the effective 1D form
\begin{equation}
\hat{H}_{1\text{D}}/\hbar = \sum_{d}(\Delta_d + J'_dL'_d)\ket{\text{W}_d}\bra{\text{W}_d}
+\sum_{d} J_dL_d\sqrt{\frac{K_d}{K_{d+1}}}\Big(\ket{\text{W}_{d+1}}\bra{\text{W}_{d}} + \text{h.c.}\Big),
\label{eqapp:Hflat}
\end{equation}
where $J_d' \coloneqq \bra{\text{q}_d^l}\hat{H}\ket{\text{q}_d^k}$ denotes the intra-layer couplings, $J_d \coloneqq \bra{\text{q}_{d+1}^l}\hat{H}\ket{\text{q}_d^k}$ the inter-layer couplings, and \mbox{$\Delta_d \coloneqq \bra{\text{q}_d^k}\hat{H}\ket{\text{q}_d^k}$} the on-site energies. 
The effective coupling strengths $J_dL_d\sqrt{K_d/K_{d+1}}$ arise from the coherent sum of all physical interactions connecting the adjacent layers, as described by the beam-splitter effect illustrated in Supplementary~Fig.~\ref{figapp:heavyhex}\textbf{b}.

We can now use the framework presented in the main text to design $\hat{H}_{1\text{D}}$ such that
\begin{equation*}
\left|\langle\text{W}_d|e^{-i\hat{H}_{1\text{D}}\tau/\hbar}|\text{W}_0\rangle\right|^2 = \frac{K_d}{N},
\end{equation*}
thereby generating a W~state across all $N \coloneqq \sum_d K_d$ qubits at time $\tau$, starting from an excitation in \mbox{$|\text{W}_0\rangle=\ket{\text{q}_0^0}$}.
In~the heavy-hex example, the couplings $J_d$ and on-site energies $\Delta_d$ satisfying this condition are presented in Supplementary~Fig.~\ref{figapp:heavyhex}\textbf{c}, while the corresponding population dynamics of the $\ket{\text{W}_d}$ states are shown in Supplementary~Fig.~\ref{figapp:heavyhex}\textbf{d}.
The states at times $0$, $\tau$ and $2\tau$ are shown in the original qubit basis in Supplementary~Fig.~\ref{figapp:heavyhex}\textbf{e}, confirming the generation of a $28$-qubit W~state at time $\tau$.

\begin{figure*}[t!]
\centering
\includegraphics{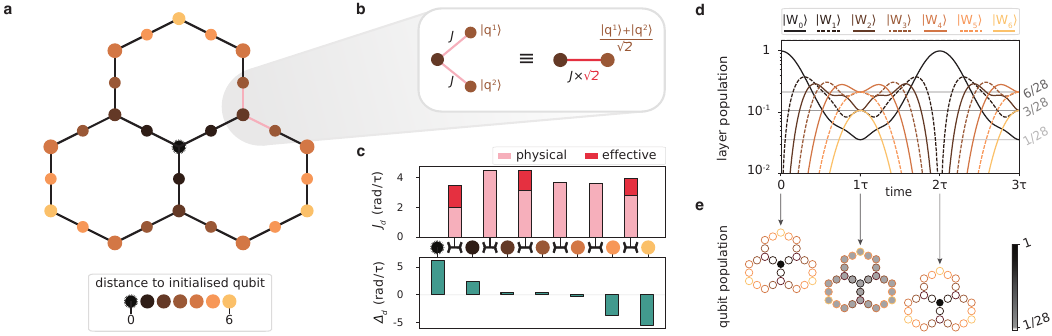}
\caption{
    \label{figapp:heavyhex}
\textbf{Beam-splitter effect and single-step entanglement in a heavy-hex lattice.}
\textbf{a} Example of a heavy-hex connectivity map comprising $N=28$ qubits (nodes).
Node colours encode the distance from the centre qubit (black), which is initialised in its excited state.
\textbf{b} Close-up of a beam-splitter structure, where one qubit is identically coupled to multiple neighbouring qubits (left~diagram), producing an effective interaction with their symmetric superposition state (right~diagram).
The number of physical connections (pink) determines the effective coupling strength (red).
\textbf{c} Required couplings $J_d$ and local energies $\Delta_d$ for exact W~state generation in the heavy-hex lattice, plotted against the distance $d$ from the centre. 
Physical couplings between qubits at distance $d$ and $d+1$ are represented in pink, while their combined effective coupling between the symmetric superposition states $\ket{\text{W}_d}$ and $\ket{\text{W}_{d+1}}$ is shown in~red.
\textbf{d} Simulated dynamics, with populations shown relative to each state $\ket{\text{W}_d}$, for $d=0,1,\ldots,6$.
At the synthesis time $\tau$, the populations are $K_d/N$, where $K_d$ denotes the number of qubits comprising $\ket{\text{W}_d}$.
\textbf{e} Individual qubit populations at times $0$, $\tau$ and $2\tau$.
}
\end{figure*}

\subsection{-- Excitation dynamics for different 1D system sizes and initial states}
\label{app_all_dynamics}
We performed operations for direct W~state generation in 1D registers of varying size, $3\leq M \leq 7$, starting from an excitation localised in a qubit close to the centre.
The population dynamics recorded in each case, including alternative choices of initialised qubits, are compared to numerical simulations in Supplementary~Fig.~\ref{figapp:alldynamics}.
Notably, for the selected $M=3$ operation, an excitation prepared on an outer qubit is partially transferred to the opposite side of the chain at regular intervals.
Indeed, this particular symmetric Hamiltonian structure has previously been proposed for implementing fractional state revival in spin chains~\cite{genest2016_fractional, Naegele2022_FST}.
By contrast, all subsequent operations with $M\geq4$ are constructed from the antisymmetric Hamiltonian family introduced in the main text (see also \ref{app_theory}), which is characterised by a linearly-spaced spectrum.
As such, the excitation refocuses to its initial state after a fixed period that is independent of the choice of initialisation.

\begin{landscape}
\begin{figure}[]
\centering
\includegraphics[width=\linewidth,
  height=0.9\textheight,
  keepaspectratio]{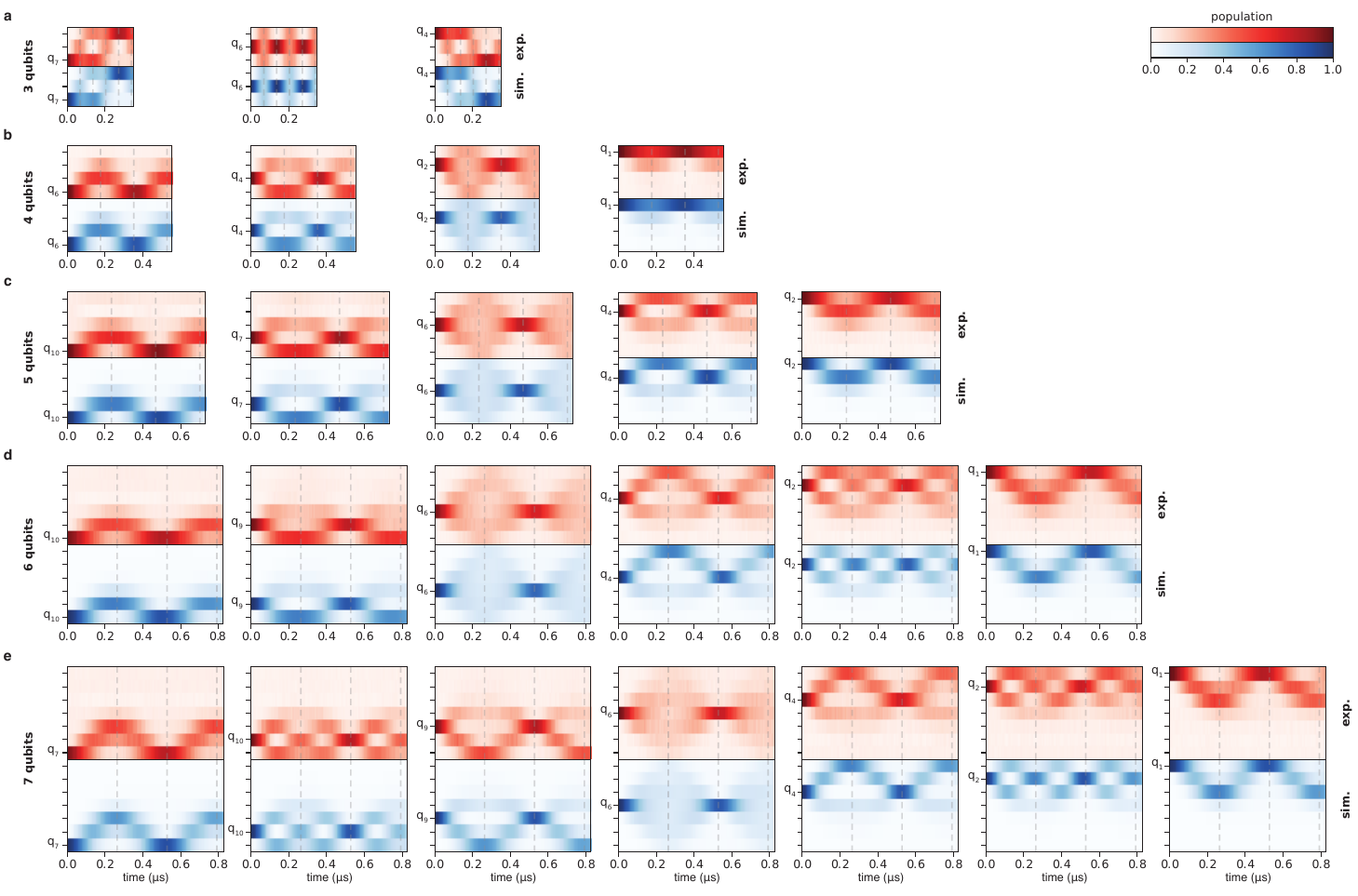}
\caption{
\textbf{Single-excitation dynamics for different 1D chain lengths and initialised qubits}. Dynamics are grouped by length (rows), ranging from $M=3$ to $7$ qubits (\textbf{a}--\textbf{e}, respectively). 
Columns correspond to different choices of initialised qubit. 
Experimental data is shown in red (top panels), while numerical simulation including dephasing and excitation loss is provided in blue (bottom panels).
Dashed lines show integer multiples of the synthesis time.
\label{figapp:alldynamics}
}
\end{figure}

\end{landscape}
\end{document}